\documentclass[12pt,a4paper]{article}
\pdfoutput=1
\usepackage{jheppub}

\usepackage{amstext,amsfonts,amsthm}

\numberwithin{equation}{section}

\usepackage[table]{xcolor}

\usepackage{verbatim}
\usepackage{bbold}
\usepackage{array}
\usepackage{graphicx}
\usepackage{multirow}
\usepackage{booktabs}
\usepackage{longtable}

\usepackage{tikz}
\usetikzlibrary{
	arrows,
	backgrounds,
	chains,
	decorations.pathmorphing,
	decorations.markings,
	er,
	fit,
	matrix,
	plothandlers,
	positioning,
	scopes,
	shapes,
	topaths
}

\makeatletter
\tikzset{join/.code=\tikzset{after node path={%
\ifx\tikzchainprevious\pgfutil@empty\else(\tikzchainprevious)%
edge[every join]#1(\tikzchaincurrent)\fi}}}
\makeatother

\tikzset{>=stealth',every on chain/.append style={join},
         every join/.style={->}}

\tikzstyle{every picture}=[baseline=(current bounding box.center)]

\tikzstyle{diag}=[->, semithick,>=stealth']

\tikzstyle{mmn}=[
	matrix of math nodes, 
	row sep=1.5cm, 
	column sep=1.5cm,
	minimum height=1.5em,
	minimum width=1.5em, 
	text height=1.5ex, 
	text depth=0.25ex]

\tikzstyle{alignTikz}=[
	minimum height=1.5em,
	minimum width=1.5em, 
	text height=1.5ex, 
	text depth=0.25ex, 
	inner sep=2pt
]

\tikzstyle{smNode}=[minimum width=0.5cm]
\tikzstyle{lNode}=[minimum width=1.3cm]

\definecolor{homCol}{rgb}{0.49,0.04,0.04}
\definecolor{SecondhomCol}{rgb}{0.12,0,0.80}
\definecolor{idCol}{rgb}{0,0,0}
\definecolor{homConeCol}{rgb}{0,0.3,0}

\tikzstyle{morph}=[->, semithick,>=stealth']

\tikzstyle{hom}=[draw=homCol,homCol,
			->, semithick,>=stealth',
			decorate,
			decoration={
				snake,
				pre length=10pt,
				post length=10pt,
				amplitude=0.5mm, 
				segment length=1.5mm
				}]
        
\tikzstyle{homSO}=[draw=SecondhomCol,
				SecondhomCol,->,  thick,>=stealth',
				decorate,decoration={snake,pre length=10pt,post length=10pt,amplitude=0.5mm,
        segment length=3mm}]

 \tikzstyle{homCone}=[draw=homConeCol,homConeCol,
 				  thick,
 				 double,-implies,double equal sign distance,
 				decorate,decoration={snake,pre length=10pt,post length=10pt,amplitude=0.5mm,
         segment length=3mm}]

 \tikzstyle{idmorph}=[draw=idCol,idCol,double,semithick,-implies,double equal sign distance]

\newcounter{planeDistance}
\newcounter{rowSep}
\newcounter{colSep}

\newcommand{\one}{\ensuremath{\mathbf{1}}}

\usepackage[colorinlistoftodos,draft]{todonotes} 

\newcommand{\id}{\text{id}}

\newcounter{NumberOfRepeatedColumns}

\hypersetup{
	colorlinks=true, 
	breaklinks=true, 
	linkcolor=green!40!black, 
	citecolor=green!40!black,
	menucolor=green!40!black, 
	urlcolor=green!40!black
	}

\usepackage{pgfplots} 

\title{Fusion of interfaces in Landau-Ginzburg models: a functorial approach}

\author[a]{Nicolas Behr}
\author[b,c]{and Stefan Fredenhagen}
 
\affiliation[a]{Universit{\'e} de Paris, CNRS, Institut de Recherche en Informatique Fondamentale\\
8 Place Aurelie Nemours, 75205 Paris Cedex 13, France
}

\affiliation[b]{University of Vienna, Faculty of Physics,\\ Boltzmanngasse 5, 1090 Vienna, Austria}
\affiliation[c]{Erwin Schr{\"o}dinger International Institute for Mathematics and Physics,\\ University of Vienna, Boltzmanngasse 9, 1090 Vienna, Austria}

\emailAdd{nicolas.behr@irif.fr}
\emailAdd{stefan.fredenhagen@univie.ac.at}

\abstract{We investigate the fusion of B-type interfaces in two-dimensional supersymmetric Landau-Ginzburg models. In particular, we propose to describe the fusion of an interface in terms of a \emph{fusion functor} that acts on the category of modules of the underlying polynomial rings of chiral superfields. This uplift of a functor on the category of matrix factorisations simplifies the actual computation of interface fusion. Besides a brief discussion of minimal models, we illustrate the power of this approach in the $SU(3)/U(2)$ Kazama-Suzuki model where we find fusion functors for a set of elementary topological defects from which all rational B-type topological defects can be generated.}

\preprint{UWThPh 2020-32} 

\begin{document}
\maketitle

\newpage
\section{Introduction}

Defects or interfaces are an important tool and constitute crucial structures in two-dimensional field theories. Topological defects which can be arbitrarily deformed as long as they do not hit other structures can be viewed as generalised symmetries~\cite{Frohlich:2004ef,Bachas:2008jd,Bachas:2012bj}, and they also open up the possibility to define generalised orbifold theories~\cite{Frohlich:2009gb,Carqueville:2012dk,Brunner:2013ota,Brunner:2013xna}. Furthermore, defects and interfaces are useful in the context of renormalisation group flows. Interfaces can encode bulk flows and separate UV and IR limits of a flow~\cite{Brunner:2007ur,Gaiotto:2012np}. One can also study flows of defects and their interplay with boundary renormalisation group flows when the defects are fused to a boundary of the theory~\cite{Graham:2003nc,Bachas:2004sy,Runkel:2007wd,Bachas:2009mc,Brunner:2009zt,Fredenhagen:2009tn}. The possibility to fuse defects and interfaces often is essential in these applications.

$N=(2,2)$ supersymmetric Landau-Ginzburg models form a basic class of two-dimensional field theories, and by flowing to the IR they provide an important construction of $N=(2,2)$ superconformal models~\cite{Vafa:1988uu}. These theories contain a topological subsector that is not affected by the flow, which gives the possibility to compute protected quantities or structures of the full superconformal theory. Moreover, boundary conditions and defects, or, more generally, interfaces between different Landau-Ginzburg models that preserve enough supersymmetry can be captured in the topological theory. There, interfaces preserving B-type supersymmetry are specified as matrix factorisations of the difference of the superpotentials of the theories that are separated by the interface~\cite{Brunner:2007qu}. Fusion of interfaces is then described by taking the tensor product of the factorisations. 

These tensor products formally lead to matrix factorisations with matrices of infinite size because they still contain the variables of the model in between the interfaces that disappeared in the process of fusion. The reduction to finite-size matrices is always possible, but in practice can be a technically cumbersome procedure~\cite{Carqueville:2011zea,Dyckerhoff:2011vf,Carqueville:2013usa}. Many interesting interfaces, however, can be cast in a form where the result of fusing it to another interface is given by an operator-like map of one finite matrix to another one. This map can be described as a functor between the categories of modules over the polynomial rings that are involved.
\smallskip

In this article we want to propose such a functorial prescription as a tool to facilitate the computation of fusion. As discussed before, a better understanding of fusion is useful for many applications of defects. Furthermore, it might simplify the analysis of the algebraic structure of the (semi-)ring of topological defects. In contrast to the fusion as a functor on the category of matrix factorisations, the \emph{fusion functors} that we consider act on the category of modules. They induce a functor on the category of matrix factorisations, and conversely can be seen as an uplift of such a functor.   

We explicitly show how such an uplift can be obtained for the topological B-type defects in minimal models. These defects can all be obtained from a set of elementary symmetry defects by means of the cone construction in the category of matrix factorisations. By defining an analogue of the cone construction for fusion functors, we show how to obtain fusion functors for all such defects.

Our main example to illustrate the use of fusion functors is the $SU(3)/U(2)$ Kazama-Suzuki model, for which matrix factorisations corresponding to rational topological B-type defects have been only known for very few examples. To obtain fusion functors in these models, we mainly use two tools: on the one hand we employ a fusion functor for an interface separating a Kazama-Suzuki model and a pair of minimal models, and on the other hand we make use of the cone construction where we are guided by known defect renormalisation group flows. We propose fusion functors for a set of elementary defects from which all rational defects can be obtained by fusion.
\smallskip

The article is organised as follows. In section~\ref{sec:IFF} we provide the necessary background on matrix factorisations in Landau-Ginzburg models and then introduce the fusion functors as well as the cone construction. Section~\ref{sec:minimalmodels} illustrates the concept in the case of minimal models. We then turn in section~\ref{sec:RDKSM} to the $SU(3)/U(2)$ Kazama-Suzuki models. After a brief review of the conformal field theory results on rational B-type defects, we discuss the fusion functor that describes an interface between these models and a pair of minimal models. By using this interface as well as the cone construction, we construct fusion functors for rational defects in the Kazama-Suzuki models and outline how to obtain all of them by successive fusion. We conclude in section~\ref{sec:conclusion}, and some technical results are to be found in two appendices.

\section{Interfaces and fusion functors}\label{sec:IFF}

B-type interfaces in $N=(2,2)$ supersymmetric Landau-Ginzburg models can
be described by matrix factorisations of the difference of the corresponding
superpotentials~\cite{Brunner:2007qu}. The fusion of interfaces is realised by the (graded) tensor
product of matrix factorisations. These tensor products still contain the variables of the theory that disappeared in the process of fusion. By eliminating these variables one can arrive at a factorisation with matrices of finite size. For many interesting interfaces this map can be realised as a simple functor between categories of modules that directly maps matrices to matrices.

In this section, after a review of B-type interfaces in Landau-Ginzburg models and their realisation as matrix factorisations, we will define such fusion functors and explain how they are related to matrix factorisations. We will also define an analogue of the cone construction for fusion functors which provides a way to construct new fusion functors from known ones.

\subsection{B-type interfaces in Landau-Ginzburg models}\label{sec:LGdefectMFs}

\begin{figure}
    \centering
    \begin{tikzpicture}[very thick,scale=2,color=blue!50!black]
\fill[green!23,rounded corners=15pt] (0,0) rectangle (3.2,2);
\draw[thick,black] (1.6,0) to (1.6,2);
\fill[color=black] (1.7,2.2) node (X) {\begin{minipage}{1cm}$Q$\end{minipage}};
\fill[color=black] (.9,1) node (X) {\begin{minipage}{1cm}$W$\end{minipage}};
\fill[color=black] (2.4,1) node (X) {\begin{minipage}{1cm}$W'$\end{minipage}};
\end{tikzpicture}
    \caption{An interface separating two Landau-Ginzburg models with superpotentials $W$ and $W'$}
    \label{fig:defect}
\end{figure}
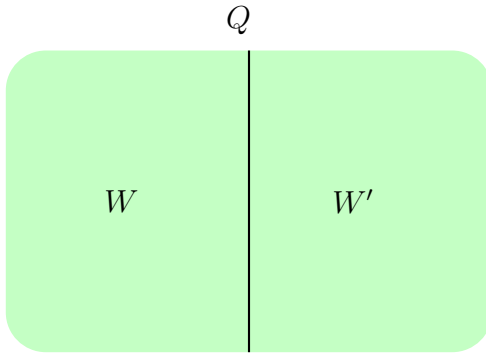

Suppose we are given two $N=(2,2)$ supersymmetric Landau-Ginzburg models
with polynomial superpotentials $W (x_{1},\dotsc ,x_{n})$ and $W'
(x_{1}',\dotsc ,x_{n'}')$ where $x_i$, $x_{i'}'$ are chiral scalar superfields. The superpotentials are thus elements of
the polynomial rings $R=\mathbb{C}[x_{1},\dotsc ,x_{n}]$ and
$R'=\mathbb{C}[x_{1}',\dotsc ,x_{n'}']$, respectively. An interface separating these theories (see figure~\ref{fig:defect}) generically breaks supersymmetry, but by introducing fermion fields on the interface with a suitable potential one can preserve a B-type combination of supersymmetries. Such a construction is possible whenever one finds a matrix factorisation of the difference of the superpotential~\cite{Brunner:2007qu} defined as follows. Given a
$\mathbb{Z}_{2}$-graded free $(R,R')$-bimodule $M=M_{0}\oplus
M_{1}$, we call an odd endomorphism $Q$ on $M$ (mapping $M_{0}$ to
$M_{1}$ and vice versa) a matrix factorisation of the difference of the
superpotentials $W$ and $W'$ if 
\begin{equation}
Q^{2}= (W-W')\cdot \id\ .
\end{equation}
If the free module $M$ has finite rank $2r$, we can realise $Q$ by an
$2r\times 2r$ matrix with polynomial entries in the variables
$x_{1},\dotsc ,x_{n}$ and $x_{1}',\dotsc ,x_{n'}'$, 
\begin{equation}
Q= \begin{pmatrix}
0 & Q^{(1)}\\
Q^{(0)} & 0 
\end{pmatrix} =
\begin{pmatrix}
0 & \dotsb & 0 & Q^{(1)}_{11} & \dotsb & Q^{(1)}_{1r}\\
\vdots &\ddots & \vdots & \vdots & \ddots& \vdots \\
0 & \dotsb & 0 & Q^{(1)}_{r1} & \dotsb & Q^{(1)}_{rr}\\
Q^{(0)}_{11}& \dotsb & Q^{(0)}_{1r} & 0 & \dotsb & 0 \\
\vdots & \ddots & \vdots &\vdots &\ddots & \vdots \\
Q^{(0)}_{r1} & \dotsb & Q^{(0)}_{rr} & 0 & \dotsb & 0 
\end{pmatrix}\ .
\end{equation}
A special situation occurs when one of the sides describes a trivial
theory, e.g.\ $W'=0$ and $R'=\mathbb{C}$. The matrix factorisations
$Q$ of the superpotential $W$, $Q^{2}=W\cdot \one$, then
describe B-type boundary conditions in the corresponding
Landau-Ginzburg model~\cite{Kontsevich:unpublished,Kapustin:2002bi,Orlov:2003yp,Brunner:2003dc,Kapustin:2003ga}.

Going back to the general situation of interfaces between non-trivial
theories, we now want to consider their spectra.

The interface fields between two interfaces described by matrix
factorisations $P$ (with underlying module $N$) and $Q$ (with
underlying module $M$) are described by a cohomology group of module
homomorphisms $\phi :N\to M$ built as follows: we define the operator
$\delta_{P,Q}$ that on a homomorphism $\phi$ of $\mathbb{Z}_{2}$-degree
$|\phi|$ acts as
\begin{equation}
\delta_{P,Q} (\phi) = Q\, \phi - (-1)^{|\phi|}\phi\, P  \ .
\end{equation}
The interface fields then correspond to $\delta_{P,Q}$-closed
homomorphisms modulo $\delta_{P,Q}$-exact homomorphisms. 

\begin{figure}
    \centering
    \begin{tikzpicture}[very thick,scale=2,color=blue!50!black]
\fill[green!23,rounded corners=15pt] (0,0) rectangle (3.2,2);
\draw[thick,black] (1,0) to (1,2);
\draw[thick,black] (2.2,0) to (2.2,2);
\fill[color=black] (1.1,2.2) node (X) {\begin{minipage}{1cm}$P$\end{minipage}};
\fill[color=black] (2.3,2.2) node (X) {\begin{minipage}{1cm}$Q$\end{minipage}};
\fill[color=black] (.6,1) node (X) {\begin{minipage}{1cm}$W$\end{minipage}};
\fill[color=black] (1.7,1) node (X) {\begin{minipage}{1cm}$W'$\end{minipage}};
\fill[color=black] (2.7,1) node (X) {\begin{minipage}{1cm}$W''$\end{minipage}};
\fill[green!23,rounded corners=15pt] (4,0) rectangle (7.2,2);
\draw[thick,black] (5.6,0) to (5.6,2);
\fill[color=black] (5.7,2.2) node (X) {\begin{minipage}{1cm}$P\otimes Q$\end{minipage}};
\fill[color=black] (4.9,1) node (X) {\begin{minipage}{1cm}$W$\end{minipage}};
\fill[color=black] (6.4,1) node (X) {\begin{minipage}{1cm}$W''$\end{minipage}};
\end{tikzpicture}
    \caption{The fusion of the interfaces described by $P$ and $Q$ leads to a new interface described by the tensor product $P\otimes Q$. The theory with superpotential $W'$ disappears in this process.}
    \label{fig:defectfusion}
\end{figure}
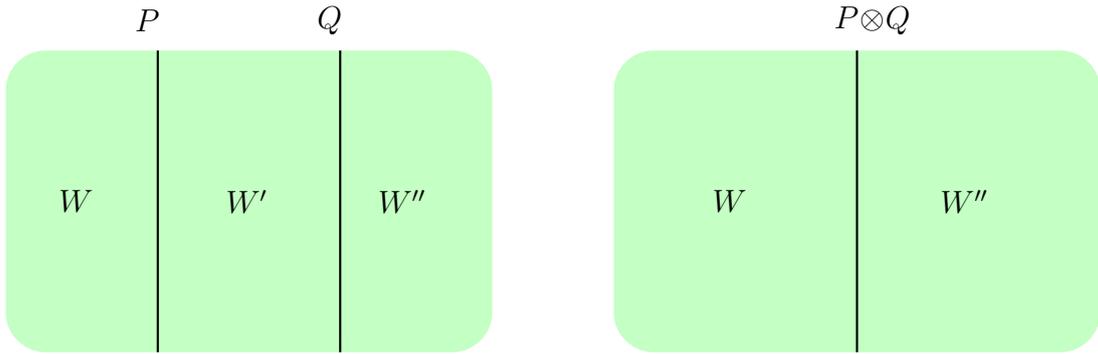

Given two parallel interfaces that separate three Landau-Ginzburg theories with
superpotentials $W$, $W'$, and $W''$, we can fuse them (see figure~\ref{fig:defectfusion})
to produce a new interface between theories with superpotentials $W$ and
$W''$~\cite{Petkova:2000ip,Brunner:2007qu}. The mathematical operation
that realises this fusion is the graded tensor product of matrix
factorisations~\cite{Yoshino:1998,Khovanov:2004,Ashok:2004zb}. Given a
$(W,W')$-interface $P$ and a $(W',W'')$-interface $Q$ the tensor
product $P\otimes Q$ is defined as the matrix
\begin{equation}
P\otimes Q = \begin{pmatrix}
 0 & 0 & P^{(1)}\otimes_{R'} \one & \one \otimes_{R'} Q^{(1)}\\
 0 & 0 & -\one\otimes_{R'}Q^{(0)} & P^{(0)}\otimes_{R'} \one\\
P^{(0)} \otimes_{R'}\one & -\one\otimes_{R'}Q^{(1)} & 0 & 0\\
\one\otimes_{R'}Q^{(0)} & P^{(1)}\otimes_{R'}\one & 0 & 0 
\end{pmatrix}
\end{equation}
acting on 
\begin{equation}
N \otimes_{R'} M = \big(N_{0}\otimes_{R'}M_{0} \oplus
N_{1}\otimes_{R'}M_{1} \big) \oplus \big(N_{1}\otimes_{R'}M_{0} \oplus
N_{0}\otimes_{R'}M_{1}\big) \ .
\end{equation}
The fusion of an interface $P$ also acts on the spectrum of interface
fields. Given two $(W',W'')$-interfaces $Q_{A}$ and $Q_{B}$, a
homomorphism $\phi :M_{A}\to M_{B}$ is mapped to
$\one_{N}\otimes_{R'}\phi:N\otimes_{R'}M_{A}\to
N\otimes_{R'}M_{B}$. This map is compatible with closure and exactness
conditions, so that the interface fusion has a well-defined action on
the interface fields. In fact, the fusion of an interface
factorisation $P$ defines a functor between categories of matrix
factorisations~\cite{Khovanov:2004}.

Computing the fusion is in general not an easy task. 
Even if $N$ and $M$ are finite rank $(R,R')$- and 
$(R',R'')$-bimodules, respectively, the module $N\otimes M$ is an
infinite rank $(R,R'')$-bimodule, which is reflected in the
fact that $P\otimes Q$ still contains the variables $x'$ that belonged
to the Landau-Ginzburg theory that disappeared in the process of
fusion. It can be shown~\cite{Khovanov:2004} that starting from finite rank modules
$N$ and $M$ the factorisation $P\otimes Q$ is isomorphic to a
matrix factorisation $T$ on a finite rank $(R,R'')$-bimodule $L$, in the
sense that there are even homomorphisms $\lambda:N\otimes M\to L$ and
$\lambda^{-1}: L \to N\otimes M$ that are $\delta_{P\otimes Q,T}$-closed and $\delta_{T,P\otimes Q}$-closed, respectively, satisfying
\begin{equation}
\lambda \,\lambda^{-1} = \one_{L} +
\big(\delta_{T,T}\text{-exact terms} \big)\quad , \quad 
\lambda^{-1}\, \lambda  = \one_{N\otimes M} + \big(
\delta_{P\otimes Q,P\otimes Q}\text{-exact terms} \big) \ .
\end{equation}
In practice, however, reducing the tensor product to a finite matrix
factorisation is often tedious (see e.g.\ \cite{Carqueville:2011zea}). 

There are, however, matrix factorisations that behave in a simple way under fusion. The simplest example is the identity defect. This corresponds to a $(W,W)$-interface across which all observables are continuous, so it behaves like a Landau-Ginzburg model with superpotential $W$ without any defect present. Even such a trivial situation can be modelled as a certain matrix factorisation that we want to call $I_{W}$. Its precise realisation as matrix
factorisation $I_{W}:M_{I_{W}}\to M_{I_{W}}$ can be found e.g.\
in~\cite{Khovanov:2004,Kapustin:2004df,Brunner:2007qu,Carqueville:2009ev}. For example, when we have one variable, the identity defect factorisation can be written as
\begin{equation}\label{identitydefect}
I_W = \begin{pmatrix}
0 & x-x'\\
\frac{W(x)-W(x')}{x-x'} & 0 \end{pmatrix} \ .
\end{equation}
Taking the tensor product of $I_{W}$ and a matrix factorisation $Q:M\to M$ that describes some $(W,W')$-interface results in a $(W,W')$-matrix factorisation $I_{W}\otimes Q$ realised on a module of infinite rank. But because the effect of fusing the identity defect is trivial, this matrix factorisation is isomorphic to $Q$, and concrete
isomorphisms $\lambda_{Q}$ and $\lambda_{Q}^{-1}$ relating $I_{W}\otimes Q$ and
$Q$ are known (see~\cite{Carqueville:2009ev,Carqueville:2012st}), and they satisfy
\begin{equation}
\lambda_{Q}\,\lambda_{Q}^{-1} = \one_{M} \, ,\quad \lambda_{Q}^{-1}\, \lambda_{Q} = \one_{M_{I_{W}}\otimes M} + \text{exact terms.}
\end{equation}
Also the action on interface fields between $(W,W')$-interfaces $Q_{A}$ and $Q_{B}$ is trivial: a morphism $\phi :M_{A}\to M_{B}$ between $Q_{A}$ and $Q_{B}$ is mapped first to $\one \otimes \phi$, which is then related to $\phi$ via
\begin{equation}\label{identityonmorphisms}
\lambda_{Q_{B}}\big(\one \otimes \phi  \big) \lambda_{Q_{A}}^{-1} = \phi   \ .
\end{equation}
The net result is that fusing the identity defect on a matrix factorisation $Q$ gives back the same matrix factorisation $Q$, and an interface field $\phi$ is mapped in the fusion process again to $\phi$. As one expects, the identity defect acts trivially.

In the following paragraph we shall describe a more general class of interfaces whose fusion on other interfaces can also be described in a simple way.

\subsection{Fusion functors}

We have seen above that the fusion of the identity defect can be summarised in a simple way. Matrix factorisations and their morphisms can all be described by homomorphisms on modules over polynomial rings, so we could describe the action of the identity defect by the identity functor in the category of modules.

We now propose to look more generally at interfaces whose fusion process can be described by a functor on the category of modules. Let us consider two superpotentials $W$, $W'$ in some polynomial rings $R$ and $R'$, respectively. We define a $(W,W')$-\emph{fusion functor} $U$ to be a $\mathbb{C}$-linear functor from the category of (free finite rank) $R$-modules to the category of (free finite rank) $R'$-modules satisfying
\begin{equation}\label{deffusionfunctor}
U (W\cdot \phi) = W' \cdot U (\phi )\quad \text{for all homomorphisms}\ \phi .
\end{equation}
This fusion functor $U$ maps a matrix factorisation $Q:M\to M$ of the superpotential $W\in R$ (realised on a module $M$) to a factorisation $U (Q)$ of the superpotential $W'$ realised on a module $U (M)$,
\begin{equation}
U (Q)\,U (Q) = U ( Q^{2}) = U (W\cdot \one_{M}) = W'\cdot U (\one_{M}) = W' \cdot \one_{U (M)} \ ,
\end{equation}
where in addition to the defining property~\eqref{deffusionfunctor} of a fusion functor we have only used general properties of a functor.

This functor maps $W$-factorisations to $W'$-factorisations, i.e.\ it maps boundary conditions in a LG theory with superpotential $W$ to a boundary condition in a theory with superpotential $W'$. We can also act with it on interfaces by trivially extending $U$ to a functor from the category of $(R,R'')$-modules to the category of $(R',R'')$-modules where $U$ acts trivially on the $R''$-part. In this way, a $(W,W'')$-factorisation $Q:M\to M$ on a $(R,R'')$-module $M$ is mapped to a $(W',W'')$-factorisation $U(Q)$ on a $(R',R'')$-module $U(M)$.  

We now want to interpret the effect of acting with the fusion functor $U$ as the fusion of a certain interface that is described by some $(W',W)$-matrix factorisation. If we think of $U$ as describing a $(W',W)$-interface, then fusing it to the $W$-identity defect should give back the same interface. On the other hand, the result of the fusion is described by the factorisation $U(I_{W})$ that we obtain when we act with $U$ on the identity defect factorisation $I_{W}$. Therefore the candidate for an interface whose fusion is described by $U$ is $U (I_{W})$.

Let us check that this is indeed the case: for a general $(W,W'')$-factorisation $Q:M\to M$ we have to show that 
\begin{equation}
U (I_{W}) \otimes Q \cong U (Q) \ .
\end{equation}
This follows from the isomorphism between $I_{W}\otimes Q$ and $Q$ by just acting with the functor $U$ (as $U$ only acts on the $R$-module structure we have $U (I_{W})\otimes Q \cong U (I_{W}\otimes Q)$), and we find
\begin{equation}
U (Q) = U (\lambda_{Q}) \, \big(U (I_{W}) \otimes Q \big)\, U ( \lambda_{Q}^{-1}) \ .
\end{equation}
We conclude that indeed a given fusion functor $U$ describes the fusion of the interface $U (I_{W})$.

Applying the functor $U$ to a factorisation $Q$ is typically much easier than eliminating the dummy variable from the tensor product factorisation $U (I_{W})\otimes Q$. It also has another advantage: not only the action on matrix factorisations is simple, but also the action on morphisms. As we discussed above, the fusion of $U (I_{W})$ to a morphism between factorisations $Q_{A}$ and $Q_{B}$ realised by a homomorphism $\phi_{AB}:M_{A}\to M_{B}$ results in the morphism $\one \otimes \phi_{AB}$ between $U (I_{W})\otimes Q_{A}$ and $U (I_{W})\otimes Q_{B}$. Via the isomorphisms $U (\lambda_{Q_{B}})$ and $U (\lambda_{Q_{A}}^{-1})$ this is mapped to $U (\phi_{AB})$ which can be seen by applying $U$ to the relation~\eqref{identityonmorphisms}. Whenever an interface can be realised as $U (I_{W})$ with some fusion functor $U$, we thus have a tool to efficiently describe its effect via fusion.

Let us look at some examples. The simplest situation (apart from the identity defect) is a symmetry defect that implements some automorphism $\sigma :R\to R$ which leaves the superpotential invariant, $\sigma (W)=W$. Connected to $\sigma$ there is a corresponding functor $\sigma^*$ which takes the automorphism to the level of $R$-modules. Practically speaking its action on homomorphisms that can be realised by matrices with $R$-valued entries is to act with $\sigma$ on each entry. A concrete example of this kind would be $R=\mathbb{C}[x]$, $W = x^{n}$, and 
\begin{equation}
\sigma (x) = e^{\frac{2\pi i}{n}}\, x\ .
\end{equation}
The corresponding interface $\sigma^*(I_{W})$ is then a symmetry defect of the minimal model described by the Landau-Ginzburg model with superpotential $W=x^{n}$ (see section~\ref{sec:minimalmodels}).

The preceding example can be generalised to the case of a homomorphism $\iota :R\to R'$ with $\iota (W)=W'$. The homomorphism gives rise\footnote{This construction is called \emph{extension of scalars}, see also section~\ref{sec:InterfaceMMKS}. In the context of matrix factorisations the extension of scalars and the closely related restriction of scalars have also been considered in~\cite{Dyckerhoff:2011vf,Carqueville:2011zea}} to a functor $\iota^*$ which has the effect on module homomorphism to apply $\iota$ to each entry in a matrix realisation. The corresponding interfaces were discussed in~\cite{Behr:2012xg} where we called them \emph{variable transformation interfaces} since their effect consists in replacing the variables of the Landau-Ginzburg models with superpotential $W$ by the variables of the Landau-Ginzburg model with superpotential $W'$. A non-trivial example is provided by $R=\mathbb{C}[y_{1},y_{2}]$, $R'=\mathbb{C}[x_{1},x_{2}]$ and the homomorphism defined by 
\begin{equation}
\iota (y_{1}) =x_{1} +x_{2} \quad ,\quad \iota (y_{2}) = x_{1}x_{2}\,,
\end{equation}
which we discuss in more detail in section~\ref{sec:InterfaceMMKS}. When $W'=x_{1}^{n}+x_{2}^{n}$ is the superpotential of the product of two minimal models, the superpotential $W$ with $\iota (W)=W'$ is the superpotential of a Kazama-Suzuki model.

There are also a lot of examples of fusion functors that do not come from ring homomorphisms which we will meet later. A useful tool to construct and find those is a cone construction for fusion functors that we discuss in section~\ref{sec:cone}.

\subsection{Equivalences}

Different fusion functors $U,V$ can lead to isomorphic matrix factorisations. In particular, if the fusion functors are isomorphic as functors, then the corresponding matrix factorisations are also isomorphic.

Let us explain this in some more detail. We can consider natural transformations $\Phi$ as morphisms between the functors; they associate to every $R$-module $M$ a homomorphism $\Phi(M):U(M)\to V(M)$ such that for all homomorphisms $f:M\to N$ we have
\begin{equation}\label{naturality}
V(f)\, \Phi(M) = \Phi (N)\,U(f) \  .
\end{equation}
$\Phi$ is a natural \emph{isomorphism} between $U$ and $V$ if there exists an inverse $\Phi'$ in the sense that for all $R$-modules $M$ one has
\begin{equation}
\Phi(M) \, \Phi'(M) = \one_{V(M)} \quad ,\quad \Phi'(M) \, \Phi(M) = \one_{U(M)} \ .
\end{equation}
If such a natural isomorphism $\Phi$ between $U$ and $V$ exists, then they describe the fusion of the same defect, or in other words they lead to isomorphic matrix factorisations $U(I_W)$ and $V(I_W)$. 

This can be seen as follows.  Let $M_{I}$ be the module underlying the identity defect factorisation $I_{W}$. Then $\phi=\Phi(M_{I}):U(M_{I})\to V(M_{I})$ defines a morphism from $U(I_{W})$ to $V(I_{W})$, because from the naturality property~\eqref{naturality} of $\Phi$ we see that
\begin{equation}
V(I_W)\, \phi - \phi\, U(I_W) = 0 \ . 
\end{equation}
Similarly, $\phi'=\Phi'(M_{I})$ defines a morphism from $V(I_{W})$ to $U(I_{W})$, and we have
\begin{equation}
\phi \circ \phi' = \one_{V(M_{I})} \quad ,\quad \phi' \circ \phi
= \one_{U(M_{I})} \ . 
\end{equation}
Therefore, $\phi$ and $\phi'$ realise an isomorphism between the factorisations $U(I_{W})$ and $V(I_{W})$. 

The existence of a natural isomorphism between the functors is a sufficient condition to obtain isomorphic matrix factorisations, but it is not a necessary condition: one expects that also non-isomorphic functors (in the sense above) can lead to isomorphic factorisations.

How are such natural isomorphisms realised concretely? We only consider free modules of finite rank isomorphic to $R^n$, hence a fusion functor is already specified by its action on the module $R$ and on homomorphisms from $R$ to $R$ which are realised by multiplication with a polynomial $p\in R$. The module $R$ is mapped to some free finite-rank module $U(R)\cong R^{r_U}$, and we can therefore realise $U(p)$ as a $r_U \times r_U$-matrix $U(p)$. A natural isomorphism $\Phi$ between $U$ and $V$ is then realised by some invertible $r_U \times r_U$-matrix $\phi=\Phi(R)$ with polynomial entries such that
\begin{equation}
\text{for all}\ p\in R:\quad U(p) = \phi^{-1} \,V(p)\, \phi \ .
\end{equation}
In other words, from a given functor $U$ one obtains an isomorphic functor $V$ by applying a similarity transformation on the polynomial matrices $U(p)$. To conclude, let us note that on a general polynomial $m\times n$-matrix $P=(p_{ij})$, which can be seen as a homomorphism from $M=R^m$ to $N=R^n$, the relation between $U(P)$ and $V(P)$ is then
\begin{equation}
\begin{pmatrix}
U (p_{11}) & \dotsb & U (p_{1n})\\
\vdots && \vdots \\
U (p_{m1}) & \dotsb & U (p_{mn})
\end{pmatrix} = \begin{pmatrix}
\phi^{-1} & & \\
 & \ddots & \\
& & \phi^{-1}
\end{pmatrix} \begin{pmatrix}
V(p_{11}) & \dotsb & V(p_{1n})\\
\vdots &  & \vdots \\
V(p_{m1}) & \dotsb  & V (p_{mn})
\end{pmatrix} \begin{pmatrix}
\phi & & \\
& \ddots &\\
& & \phi 
\end{pmatrix}\ ,
\end{equation}
or in short
\begin{equation}
U(P)= \Phi(R^m)^{-1}\, V(P)\, \Phi(R^n)\ ,
\end{equation}
where $\Phi(R^m)$ is a block diagonal matrix with $m$ copies of the blocks $\phi$.

\subsection{Cone construction}\label{sec:cone}

In this section we discuss an analogue of the cone construction for fusion functors.
Given two $(W,W')$-matrix factorisations $Q_{A},Q_{B}$, and a
fermionic morphism $\psi$ between them, we can construct a new matrix
factorisation 
\begin{equation}
C (Q_{A},Q_{B},\psi) = \begin{pmatrix}
Q_{A} & 0\\
\psi & Q_{B}
\end{pmatrix} \ .
\end{equation}
This is the cone construction (see e.g.\ \cite{Orlov:2003yp}). One can easily see that
$C(Q_{A},Q_{B},\psi)$ is again a $(W,W')$-matrix factorisation,
\begin{equation}
C (Q_{A},Q_{B},\psi)^{2} = (W-W')\cdot \one \ ,
\end{equation}
by using that $\psi$ is closed,
\begin{equation}
\psi Q_{A} + Q_{B} \psi = 0 \ .
\end{equation}
In physics terms the cone construction describes the endpoint of a renormalisation
group flow of the superposition of interfaces given by $Q_{A}$ and
$Q_{B}$ triggered by the interface field $\psi$. In the examples, we will make
use of this connection to obtain new factorisations as cones by
comparing to known renormalisation group flows.
\smallskip

The cone construction enables us to construct new factorisations from known ones. We now want to explore whether we can generalise this construction to fusion functors. 
For two fusion functors $U_{1}$, $U_{2}$, we can build the sum $U_{1}\oplus U_{2}$ as a fusion functor that describes the fusion of the superposition of the interfaces $U_{1}(I_{W})$ and $U_{2} (I_{W})$. It maps an $R$-module $M$ to $U_{1}(M)\oplus U_{2}(M)$ and a homomorphism $\phi_{AB}:M_{A}\to M_{B}$ to an element in $\text{Hom} \big(U_{1} (M_{A})\oplus U_{2} (M_{A}),U_{1} (M_{B})\oplus U_{2} (M_{B}) \big)$ that can be described as
\begin{equation}
\big(U_{1}\oplus U_{2} \big) \big(\phi_{AB}\big) = \begin{pmatrix}
U_{1} (\phi_{AB}) & 0\\
0 & U_{2} (\phi_{AB})
\end{pmatrix}  \ .
\end{equation}
In particular, when applied to the identity defect factorisation $I_{W}$ one obtains the factorisation 
\begin{equation}
\big(U_{1} \oplus U_{2} \big) (I_{W}) = \begin{pmatrix}
U_{1} (I_{W}) & 0\\
0 & U_{2} (I_{W})
\end{pmatrix} \ .
\end{equation}
If we want to build a functor that corresponds to a cone of $U_{1} (I_{W})$ and $U_{2} (I_{W})$, the most natural ansatz would be to look for a functor $C (U_{1},U_{2},\psi)$ that acts on modules in the same way as $U_{1}\oplus U_{2}$, but on homomorphisms as
\begin{equation}
C (U_{1},U_{2},\psi) ( \phi) = \begin{pmatrix}
U_{1} (\phi) & 0\\
\Psi (\phi) & U_{2} (\phi)
\end{pmatrix} \ .
\end{equation}
We have not yet said what $\Psi$ should be, but we can infer its properties from the requirement that $C (U_{1},U_{2},\Psi)$ is a fusion functor. Obviously, $\Psi$ should map a homomorphism $\phi_{AB} :M_{A}\to M_{B}$ to a homomorphism $\Psi (\phi_{AB}):U_{1} (M_{A})\to U_{2} (M_{B})$. This map should be $\mathbb{C}$-linear. We require $C (U_{1},U_{2},\Psi)$ to be a functor, in particular it should satisfy
\begin{equation}
C (U_{1},U_{2},\Psi) ( \phi_{BC}\circ \phi_{AB}) = C (U_{1},U_{2},\Psi) (\phi_{BC}) \circ C (U_{1},U_{2},\Psi) (\phi_{AB})\ .
\end{equation}
Evaluating this condition leads to the requirement that
\begin{equation}\label{FirstConditiononPsi}
\Psi (\phi_{BC}\circ \phi_{AB}) = \Psi (\phi_{BC})\circ  U_{1} (\phi_{AB}) + U_{2} (\phi_{BC})\circ  \Psi (\phi_{AB}) \ .
\end{equation}
In addition we have to impose the fusion functor property~\eqref{deffusionfunctor} upon $C (U_{1},U_{2},\Psi )$, which translates into the condition
\begin{equation}\label{SecondConditiononPsi}
\Psi (W\cdot \phi_{AB}) = W'\cdot \Psi (\phi_{AB}) \ .
\end{equation}
As we will explain in a subsequent publication, such a map $\Psi$ can be viewed as a \emph{degree-$1$ morphism} between $U_{1}$ and $U_{2}$. We do not want to deepen this aspect here, since for the moment it is sufficient to know that we have the possibility of building a cone functor from $U_{1}$ and $U_{2}$ whenever we find a map $\Psi$ satisfying the properties~\eqref{FirstConditiononPsi} and~\eqref{SecondConditiononPsi}. We will see examples of such maps in sections~\ref{sec:minimalmodels} and~\ref{sec:ffascones}.

Let us investigate some of the properties of $\Psi$. When we apply the conditions~\eqref{FirstConditiononPsi} and~\eqref{SecondConditiononPsi} to the identity homomorphism, we find
\begin{equation}
\Psi (\one) = 0 \quad , \quad \Psi(W\cdot\one)=0 \ .
\end{equation}
Furthermore, we can conclude from~\eqref{FirstConditiononPsi} that
$\Psi(Q)$ is a closed homomorphism from $U_{1}(Q)$ to $U_{2}(Q)$,
\begin{equation}
\Psi(Q)U_{1}(Q) + U_{2}(Q)\Psi(Q) = \Psi(Q^{2}) = \Psi(W\cdot\one) = 0
\ .
\end{equation}
The cone of $U_{1} (Q)$ and $U_{2} (Q)$ built via this homomorphism $\Psi(Q)$ is the factorisation that one obtains by applying the functor $C (U_{1},U_{2},\Psi )$ to the factorisation $Q$,
\begin{equation}
C (U_{1},U_{2},\Psi ) ( Q ) = \begin{pmatrix}
U_{1} (Q) & 0\\
\Psi (Q) & U_{2} (Q)
\end{pmatrix} = C (U_{1} (Q),U_{2} (Q),\Psi (Q)) \ .
\end{equation}
In particular, the interface whose fusion is described by the functor $C (U_{1},U_{2},\Psi )$ is a cone of the factorisations $U_{1} (I_{W})$ and $U_{2} (I_{W})$,
\begin{equation}
C (U_{1},U_{2},\Psi ) (I_{W}) =  C (U_{1} (I_{W}),U_{2} (I_{W}),\Psi (I_{W})) \ ,
\end{equation}
The converse question whether every cone of $U_{1} (I_{W})$ and $U_{2} (I_{W})$ has a description as a fusion functor will be explored elsewhere. 
\smallskip

This concludes our presentation of the basic setup. In the following, we will demonstrate that fusion functors provide a powerful tool to describe fusion of interfaces. After briefly exemplifying the concept in the case of minimal models, we will apply the formalism in the $SU(3)/U(2)$ Kazama-Suzuki models, where we will show how it can be used to find fusion functors (and hence matrix factorisations) for all rational topological defects.

\section{Fusion of defects in minimal models}\label{sec:minimalmodels}

Before we move to the Kazama-Suzuki models, we apply the concept of fusion functors and their cones in the example of minimal models. This on the one hand will illustrate the basic concepts including the cone construction, and on the other hand some of the results prepare the ground for similar constructions in the Kazama-Suzuki models.
\smallskip

The $N=(2,2)$ superconformal minimal models constitute the infrared fixed-points of Landau-Ginzburg theories with a superpotential $W=x^{n}$ \cite{Vafa:1988uu}. The matrix factorisations that describe the rational B-type boundary conditions have been identified in~\cite{Brunner:2003dc,Kapustin:2003rc}, and they are given by
\begin{equation}\label{defofQl}
Q_{l} = \begin{pmatrix}
0 & x^{l+1}\\
x^{n-l-1} & 0
\end{pmatrix} \quad , \ 0\leq l \leq n-2\ .
\end{equation}
On the other hand, B-type defects in such models are described by matrix factorisations of the difference of the superpotentials~\cite{Brunner:2007qu}
\begin{equation}
x^{n}-x'^{n} = \prod_{\xi^{n}=1} (\xi x- x') \ .
\end{equation}
There exist the obvious factorisations
\begin{equation}\label{defofminmoddefects}
D_{\{\xi_{1},\dotsc ,\xi_{l} \}} = \begin{pmatrix}
0 & (\xi_{1}x-x')\dotsb (\xi_{l}x-x')\\
\prod_{\xi \not= \xi_{i}} (\xi x- x') & 0 
\end{pmatrix} \ ,
\end{equation}
labelled by sets of $n^{\text{th}}$ roots of $1$. Now we would like to
understand the effect of fusing such defects. The first observation is
that the defects $D_{\{\xi \}}$ labelled by a single root $\xi$ are symmetry defects corresponding to
the automorphisms $\sigma_{\xi}$ that act on $x$ as $x\mapsto \xi
\cdot x$ (see e.g.\ \cite{Brunner:2007qu}). We denote the corresponding fusion functors by 
$\sigma_{\xi}^*$. In particular $D_{\{1\}}=I_{W}$ is the
identity defect, $\sigma_1^*=\id$. 

Also other defects can be described by fusion functors. To begin with, we note that the defects corresponding to the other defects in~\eqref{defofminmoddefects} can be obtained as cones from the symmetry defects~\cite{Brunner:2005fv}. In particular,  
from two factorisations $D_{\{\xi_{1}\}}$,
$D_{\{\xi_{2}\}}$, one can obtain the factorisation $D_{\{\xi_{1},\xi_{2}\}}$ as the
cone
\begin{equation}
D_{\{\xi_{1},\xi_{2}\}} \cong C(D_{\{\xi_{1}\}},D_{\{\xi_{2}\}},\psi)
\end{equation} 
based on the morphism 
\begin{equation}\label{defpsi}
\psi = \begin{pmatrix}
0 & 1\\
-\frac{W-W'}{(\xi_{1}x-x') (\xi_{2}x-x')} & 0
\end{pmatrix}\ .
\end{equation}
To describe this defect as a fusion functor, we want to perform a corresponding cone construction for the fusion functors $\sigma_{\xi_{1}}^*$ and $\sigma_{\xi_{2}}^*$. This is indeed possible by defining\footnote{Note that despite the factor of $x$ appearing in the denominator, $\Psi$ applied to a polynomial will again produce a polynomial.}
\begin{equation}\label{defPsi}
\Psi = \frac{1}{x} (\sigma_{\xi_{1}}^*-\sigma_{\xi_{2}}^*) \ ,
\end{equation}
which acts on homomorphisms $\phi_{AB}:M_{A}\to M_{B}$ of $R$-modules. It satisfies
the property~\eqref{FirstConditiononPsi} of a degree-1 morphism,
\begin{align}
&\sigma_{\xi_{2}}^*(\phi_{BC})\Psi(\phi_{AB})+\Psi(\phi_{BC})\sigma_{\xi_{1}}^*(\phi_{AB}) \nonumber \\
&=
\sigma_{\xi_{2}}^*(\phi_{BC})\frac{1}{x}\big(\sigma_{\xi_{1}}^*(\phi_{AB}) - \sigma_{\xi_{2}}^*(\phi_{AB})
\big) + \frac{1}{x} \big( \sigma_{\xi_{1}}^*(\phi_{BC})-\sigma_{\xi_{2}}^*(\phi_{BC})\big) \sigma_{\xi_{1}}^*(\phi_{AB}) 
\nonumber\\
& = \frac{1}{x} \big(\sigma_{\xi_{1}}^* (\phi_{BC}\phi_{AB}) - \sigma_{\xi_{2}}^* (\phi_{BC}\phi_{AB}) \big)\nonumber\\
& = \Psi (\phi_{BC}\phi_{AB}) \ .
\end{align}
As $\sigma_{\xi}$ is a symmetry of the superpotential $W$, $\Psi$
also satisfies the second condition~\eqref{SecondConditiononPsi},
\begin{equation}
\Psi (W\cdot \phi_{AB}) = \frac{1}{x} \big(\sigma_{\xi_{1}}^* (W\cdot \phi_{AB})
-\sigma_{\xi_{2}}^* (W\cdot \phi_{AB}) \big) \cdot\one = W\cdot \Psi (\phi_{AB}) \ .
\end{equation}
Hence we can define the fusion functor
\begin{equation}\label{minmodconefunctor}
C (\sigma_{\xi_{1}}^*,\sigma_{\xi_{2}}^*,\Psi) = \begin{pmatrix}
\sigma_{\xi_{1}}^* & 0\\
\frac{1}{x} (\sigma_{\xi_{1}}^*-\sigma_{\xi_{2}}^*) & \sigma_{\xi_{2}}^*
\end{pmatrix} \ .
\end{equation}
Applying this functor to the identity matrix factorisation $I_W$ (given in~\eqref{identitydefect}) we obtain the cone
\begin{equation}
C (\sigma_{\xi_{1}}^*,\sigma_{\xi_{2}}^*,\Psi) (I_W)=
C(D_{\{\xi_{1}\}},D_{\{\xi_{2}\}},\Psi (I_{W}))
\end{equation}
with the morphism
\begin{equation}
\Psi (I_{W}) = (\xi_{1}-\xi_{2}) \begin{pmatrix}
0 & 1\\
-\frac{W(x)-W(x')}{(\xi_{1}x-x')(\xi_{2}x-x')} & 0
\end{pmatrix} 
= (\xi_{1}-\xi_{2})\psi \ ,
\end{equation}
which up to a constant factor coincides with the morphism $\psi$
(see~\eqref{defpsi}) that was used in the cone construction of
$D_{\{\xi_{1},\xi_{2}\}}$. We therefore conclude that
$D_{\{\xi_{1},\xi_{2}\}}$ can be represented by the fusion functor
$C (\sigma_{\xi_{1}}^*,\sigma_{\xi_{2}}^*,\Psi)$ given
in~\eqref{minmodconefunctor}. 

As an application, we can now easily determine the action of
the defect $D_{\{\xi_{1},\xi_{2}\}}$ on boundary conditions $Q_{l}$:
\begin{align}
& D_{\{\xi_{1},\xi_{2}\}}\otimes Q_{l}  \cong 
\begin{pmatrix}
\sigma_{\xi_{1}}^* & 0\\
\frac{1}{x} (\sigma_{\xi_{1}}^*-\sigma_{\xi_{2}}^*) & \sigma_{\xi_{2}}^* 
\end{pmatrix} \big(Q_{l} \big) \nonumber\\
&= \begin{pmatrix}
 0 & 0 & \sigma_{\xi_{1}}^* (x^{l+1}) & 0 \\
 0 & 0 & \frac{1}{x}\big(\sigma_{\xi_{1}}^* (x^{l+1})-\sigma_{\xi_{2}}^* (x^{l+1})\big) & \sigma_{\xi_{2}}^* (x^{l+1}) \\
 \sigma_{\xi_{1}}^* (x^{n-l+1})  & 0 & 0 & 0\\
 \frac{1}{x}\big(\sigma_{\xi_{1}}^* (x^{n-l+1}) - \sigma_{\xi_{2}}^*(x^{n-l+1})\big) 
& \sigma_{\xi_{2}}^* (x^{n-l+1}) & 0 & 0 
\end{pmatrix} \ .
\end{align}
If the boundary label $l$ and the roots $\xi_{1},\xi_{2}$ are such
that $\xi_{1}^{l+1}=\xi_{2}^{l+1}$, then the induced morphisms are
zero, and we are left with a superposition of two copies of
$Q_{l}$. In the generic case ($\xi_{1}^{l+1}\not=\xi_{2}^{l+1}$), we
can use a similarity transformation to bring the fusion product to the
form
\begin{equation}\label{minmodfusionresult}
D_{\{\xi_{1},\xi_{2}\}}\otimes Q_{l} \cong 
\begin{pmatrix}
0 & 0 & x^{l+2} & 0 \\
0 & 0 & 0 & x^{l}\\
x^{n-l} & 0 & 0 & 0 \\
0 & x^{n-l+2} & 0 & 0 
\end{pmatrix}\ ,
\end{equation}
i.e.\ the fusion product is a superposition of boundary conditions
$Q_{l-1}$ and $Q_{l+1}$ (if $l=0$ then $Q_{l-1}$ is the trivial
factorisation), and we reproduce the results of~\cite{Brunner:2007qu}.
\smallskip

One could now go on and compose the identified fusion functors to obtain new ones. In particular, factorisations for all rational topological defects can be obtained by taking tensor products of defect factorisations of the type $D_{\{\xi_1,\xi_2\}}$ that we just discussed~\cite{Brunner:2007qu}. In a similar way one can construct fusion functors for all rational topological defects by composing the corresponding fusion functors. We will stop our discussion of the minimal models now and move on to the Kazama-Suzuki models, where only few results on matrix factorisations for defects exist. As we will see the fusion functor formalism will allow us to construct fusion functors and matrix factorisations for all rational defects in the $SU(3)/U(2)$ Kazama-Suzuki model.

\section{Rational defects in Kazama-Suzuki models}\label{sec:RDKSM}

In this section we will discuss rational topological defects in the $SU(3)/U(2)$ Kazama-Suzuki model and their fusion in terms of fusion functors. We start by a brief summary of the conformal field theory background and introduce the rational boundary states and rational topological defects. We then introduce an interface between the Kazama-Suzuki model and a product of two minimal models which we realise by a fusion functor. This interface is used to construct a first non-trivial example of a fusion functor for a rational defect in the Kazama-Suzuki models. We then show how one can obtain fusion functors for further defects by the cone construction, and finally we discuss how this leads to a construction of fusion functors for all rational topological defects in the model.

\subsection{Conformal field theory}\label{sec:CFT}

The Kazama-Suzuki models \cite{Kazama:1988uz,Kazama:1989qp} are rational conformal field theories with $N=(2,2)$ supersymmetry which can be obtained from a coset construction. Here we concentrate on the model based on the $SU (3)/U (2)$ cosets. The spectrum can be decomposed into representations of the chiral symmetry algebra, a W-algebra extending the superconformal algebra. The representations of the bosonic subalgebra of the symmetry algebra of the coset model $SU (3)/U (2)$ at level $k$ are labelled by tuples $(\Lambda ,\Sigma ;\lambda ,\mu )$ where $\Lambda= (\Lambda_{1},\Lambda_{2})$ labels a highest-weight representation of $SU (3)$ ($\Lambda_{i}$ being the Dynkin labels, restricted by $\Lambda_{1}+\Lambda_{2}\leq k$), $\Sigma \in\{0,v,s,\bar{s}\}$ labels the trivial, vector, spinor or conjugate spinor representation of $so(4)$, $\lambda\in\{0,\dotsc ,k+1 \}$ denotes a highest weight of $SU (2)$ and $\mu$ is a $6(k+3)$-periodic integer that labels representations of $U(1)$. More details on the construction can e.g.\ be found in~\cite{Behr:2010ug}. The labelling comes from the way how these representations are constructed: the representation of the affine Lie algebra $\widehat{su (3)}\times \widehat{so (4)}$ at level $k$ and $1$, respectively, labelled by $(\Lambda ,\Sigma)$, are decomposed with respect to $\widehat{su (2)}\times \widehat{u (1)}$, and the subspace transforming in the representation labelled by $(\lambda ,\mu)$ defines a representation of the coset algebra.

Because not all representations of $\widehat{su (2)}\times \widehat{u (1)}$ appear in the decomposition, the tuples are subject to selection rules, which are given by
\begin{equation}
\frac{2\Lambda_{1}+\Lambda_{2}}{3} + \frac{|\Sigma |}{2} -\frac{\lambda}{2} + \frac{\mu}{6} \in \mathbb{Z}\ .
\end{equation}
Here, $|\Sigma |=0$ for $\Sigma =0,v$ and $|\Sigma |=1$ for $\Sigma =s,\bar{s}$. On the other hand, some of the representations of the coset algebra that are constructed in this way are equivalent, therefore some of the tuples have to be identified,
\begin{equation}\label{identrules}
\big((\Lambda_{1},\Lambda_{2}),\Sigma ;\lambda ,\mu \big) \sim \big((k-\Lambda_{1}-\Lambda_{2},\Lambda_{1}),v\times \Sigma ;k+1-\lambda ,\mu + (k+3)\big)\ .
\end{equation}
Here we denote by $v\times \Sigma$ the fusion of the vector representation with $\Sigma$ which acts as 
\begin{equation}
v\times 0 \cong v \quad ,\quad v\times v \cong 0 \quad ,\quad v\times s \cong \bar{s} \quad ,\quad v\times \bar{s} \cong s \ .
\end{equation}
We consider the model with \emph{diagonal spectrum}, where the bulk fields transform in the same representation under the left-moving and right-moving chiral symmetry algebra, and all coset primary fields appear once.

B-type boundary conditions can be obtained by following the Cardy construction~\cite{Cardy:1989ir} of twisted boundary conditions in coset models~\cite{Maldacena:2001ky,Ishikawa:2001zu,Ishikawa:2002wx,Fredenhagen:2003xf}. Concrete expressions for the model at hand have been obtained in~\cite{Ishikawa:2003kk,Behr:2010ug}. 
The corresponding boundary states $|L,S;\ell\rangle$ are labelled by integers $L$ (with $0\leq L\leq \lfloor\frac{k}{2} \rfloor $) and $\ell$ (with $0\leq \ell \leq k+1$), as well as one of four $so (4)$ representations $S\in\{0,v,s,\bar{s} \}$. We can restrict to $S\in\{0,v \}$; this corresponds to a specific sign choice in the gluing condition for the supercurrents. For the boundary state labels, we also have identification rules,
\begin{equation}
|L,0;\ell \rangle = |L,v;k+1-\ell \rangle \ . 
\end{equation}
The construction of rational topological defects, i.e.\ defects which are transparent to the energy-momentum tensor and at which all fields in the coset algebra are glued with the help of an automorphism, is well understood~\cite{Petkova:2000ip}. The B-type defects are labelled by the same labels as the bulk fields, and we denote them by
\begin{equation}
D_{(\Lambda ,\Sigma ;\lambda ,\mu)} \ .
\end{equation}
By again fixing the sign in the gluing condition for the supercurrents, we can restrict to $\Sigma\in \{0,v \}$ which we will do from now on.

A topological defect can be fused to another defect or to a boundary~\cite{Petkova:2000ip,Graham:2003nc}. When we fuse it to a boundary, we obtain
\begin{equation}\label{Actionofdefectonbs}
D_{(\Lambda ,\Sigma ;\lambda ,\mu)} |L,S;\ell \rangle = \sum_{L',\ell '} n_{\Lambda L}{}^{L'} \,N^{(k+1)}_{\lambda \,\ell}{}^{\ell '} |L',\Sigma \times S;\ell '\rangle \ , 
\end{equation}
where we think of $D_{(\Lambda ,\Sigma ;\lambda ,\mu)}$ as an operator on the bulk Hilbert space of the conformal field theory, and fusing the defect to the boundary corresponds to applying the operator to the boundary state. The constants that appear here are the fusion rules $N^{(k+1)}_{\lambda \,\ell}{}^{\ell '}$ of $\widehat{su (2)}$ at level $k+1$ (which can be found e.g.\ in~\cite[Chapter 16]{FrancescoCFT}) and the twisted fusion rules $n_{\Lambda L}{}^{L'}$ of $\widehat{su (3)}$ at level $k$ (explicit expressions in terms of fusion rules of $su(2)$ can be found in~\cite{Gaberdiel:2002qa,Behr:2010ug}). 

The fusion of rational defects among themselves can also be explicitly described, and it is given by
\begin{equation}\label{rationalfusion}
D_{(\Lambda ,\Sigma ;\lambda ,\mu )} \, D_{(\Lambda ',\Sigma ';\lambda ',\mu ')} = \sum_{\Lambda '',\lambda ''} 
N^{(k)}_{\Lambda \,\Lambda '}{}^{\Lambda ''} N^{(k+1)}_{\lambda \,\lambda '}{}^{\lambda ''}\, D_{(\Lambda '',\Sigma \times \Sigma '; \lambda '',\mu +\mu ')} \ ,
\end{equation}
where $N^{(k)}_{\Lambda \,\Lambda '}{}^{\Lambda ''}$ denotes the fusion rules of $\widehat{su (3)}$ at level $k$ (algorithms to compute such fusion rules are given e.g.\ in~\cite[Chapter 16]{FrancescoCFT}). A particular role is played by the defects $D_{((0,0),0;0,6m)}$ whose effect on any other defect is just to change its $\mu$-label,
\begin{equation}
D_{((0,0),0;0,6m)}\, D_{(\Lambda ,\Sigma ;\lambda ,\mu)} = D_{(\Lambda ,\Sigma ;\lambda ,\mu +6m)} \ .
\end{equation}
The action of such a defect is therefore invertible, and it is called a group-like or symmetry defect. Under fusion these defects form the group $\mathbb{Z}_{k+3}\equiv \mathbb{Z}/ (k+3)\mathbb{Z}$. 

We can consider superpositions of defects, and if one also formally allows for subtracting defects, they form a ring which is isomorphic to the fusion ring of the coset algebra. This ring is generated by the elements 
\begin{equation}\label{setofgeneratingdefects}
D_{((0,0),0;0,6)}\quad ,\quad D_{((0,0),0;1,3)}\quad ,\quad D_{((0,1),0;0,2)}\quad,\quad D_{((1,0),0;0,2)} \ .
\end{equation}
This can be understood as follows: up to selection and identification rules, the fusion ring essentially has the structure of the product of the fusion rings for the constituents $su (3)$, $so (4)$, $su (2)$ and $u (1)$. These fusion rings are all generated by the fundamental representations, and hence if we consider the corresponding defects in the coset model, we can generate all others by fusion and possibly subtraction of defects. In this way, we obtain all defects $D_{(\Lambda,\Sigma ;\lambda ,\mu)}$ with $\Sigma =0$, and by the identification rule~\eqref{identrules} also all those with $\Sigma =v$. When we later construct matrix factorisations and fusion functors that correspond to the rational defects, the strategy is to first construct those for the generating defects in~\eqref{setofgeneratingdefects}. All others can then in principle be obtained by fusion and subtraction.

There is one last field theory ingredient that we will need later in our constructions, namely renormalisation group flows that connect different configurations of defects. If we know that certain defects can be obtained by perturbing a given configuration of defects, we can use this information to obtain the corresponding matrix factorisation by a cone construction that describes the end point of the renormalisation group flow. There is a particular class of renormalisation group flows of boundary conditions or defects in coset models that can be summarised in a simple rule \cite{Fredenhagen:2002qn,Fredenhagen:2003xf,Bachas:2009mc}. In our case, this rule predicts the defect flows
\begin{equation}\label{eq:FSflowsDefDef}
\sum_{\gamma ,\nu} b^{\Xi^{+}}_{(\gamma;\nu)}\,N^{(k+1)}_{\gamma
\lambda}{}^{\lambda'}\, D_{(\Lambda,\Sigma;\lambda',\mu+\nu)} 
\ \longrightarrow \ \sum_{\Lambda'}
N^{(k)}_{\Xi\Lambda}{}^{\Lambda'}\,
D_{(\Lambda',\Sigma;\lambda,\mu)} \ .
\end{equation}
Here $\Xi=(\Xi_{1},\Xi_{2})$ are the Dynkin labels of a unitary
representation of $\widehat{su(3)}_{k}$. The branching coefficients
$b^{\Xi^+}_{(\gamma;\nu)}$ indicate how often the representation labelled by $(\gamma;\nu)$ of $su(2)\times u(1)$ appears in the decomposition of the $su(3)$-representation $\Xi^+$ conjugate to $\Xi$,
\begin{equation}
\Xi^+ = (\Xi_2,\Xi_1) \to \sum_\gamma b^{\Xi^+}_{(\gamma;\nu)} \, (\gamma;\nu) = \sum_{\alpha_1=0}^{\Xi_2} \sum_{\alpha_2=0}^{\Xi_1}  \big(\alpha_1 + \alpha_2 ; 3(\alpha_1 -\alpha_2) + 2(\Xi_1 -\Xi_2)\big) \ .
\end{equation}

A simple example of a flow is obtained by setting $\Xi=(1,0)$, $\Lambda=
(0,0)$, $\Sigma=0$, $\lambda=0$, and $\mu=4$:
\begin{equation}\label{defectflowI}
D_{( (0,0),0;1,3)} + D_{( (0,0),0;0,6)} \ \longrightarrow \ 
D_{( (1,0),0;0,4)} \ . 
\end{equation}
Further examples of flows are
\begin{align}\label{eq:flowD01}
D_{((0,0),0;1,3)} + D_{((0,0),0;0,0)} \ &\longrightarrow 
\ D_{((0,1),0;0,2)} \\
D_{((0,0),0;2,6)} + D_{((0,0),0;1,3)} + D_{((0,0),0;0,0)} 
\ &\longrightarrow \ D_{((0,2),0;0,4)}\label{eq:flowD02}\\
D_{((0,0),0;2,6)} + D_{((0,0),0;1,9)} + D_{((0,0),0;0,12)}
\ &\longrightarrow \ D_{((2,0),0;0,8)}\label{eq:flowD20}\\
D_{((0,0),0;2,6)}+ D_{((0,0),0;1,9)} + D_{((0,0),0;1,3)}
+ D_{((0,0),0;0,6)} \ &\longrightarrow \ D_{((1,1),0;0,6)} \ . \label{eq:flowD11}
\end{align}
This shows that one can obtain all B-type defects by flows from those
with trivial $su(3)$ labels. In particular -- by combining flows and fusion -- one can generate all defects just from two elementary ones,\footnote{Note that by successively fusing $D_{((0,0),0;0,6)}$ we can always shift the $\mu$-label to any allowed value.} $D_{((0,0),0;0,6)}$ and $D_{((0,0),0;1,3)}$.

This concludes our brief summary of the conformal field theory construction of the $SU (3)/U (2)$ Kazama-Suzuki model. More details can be found e.g.\ in~\cite{Behr:2010ug,Behr:2012xg}.

\subsection{Interface between minimal models and Kazama-Suzuki models}\label{sec:InterfaceMMKS}

The $SU (3)/U (2)$ Kazama-Suzuki model can be obtained as the infrared fixed-point of a Landau-Ginzburg model with superpotential \cite{Lerche:1989uy,Gepner:1988wi}
\begin{equation}\label{eq:Wsu3u2}
W (y_{1},y_{2}) = \sum_{i=0}^{\lfloor \frac{k+3}{2}\rfloor}
y_{1}^{k+3-2i} y_{2}^{i} \, (-1)^{i} \, \frac{k+3}{k+3-i} \binom{k+3-i}{i} \ .
\end{equation}
The form of the potential becomes particularly simple when we express the variables $y_{i}$ in terms of variables $x_{i}$ with the substitution
\begin{equation}
y_{1} \mapsto Y_{1} (x_{1},x_{2}) = x_{1}+x_{2} \quad ,\quad y_{2}\mapsto Y_{2} (x_{1},x_{2}) = x_{1}x_{2} \ ,
\end{equation}
i.e.\ we think of the $y_{i}$ as the elementary symmetric polynomials in the variables $x_{i}$. We then have
\begin{equation}
W (Y_{1} (x_{1},x_{2}),Y_{2} (x_{1},x_{2})) = x_{1}^{k+3} + x_{2}^{k+3} = W' (x_{1},x_{2})\ ,
\end{equation}
which is the superpotential of the product of two minimal models.

This simple relation between the superpotential of a Kazama-Suzuki model and of two minimal models leads to the construction of an interface between these different theories as we explain in a moment. Because the central charges of the corresponding two conformal field theories are different, this interface cannot be a topological interface in the conformal field theory, but it will still describe a conformal interface. 

The construction of the interface proceeds via the fusion functor formalism. The maps $Y_{i}$ introduced above lead to a ring homomorphism
\begin{equation}\label{defofiota}
\iota : \mathbb{C}[y_{1},y_{2}] \ \to \ \mathbb{C}[x_{1},x_{2}] 
\end{equation}
with $\iota (W)=W'$. This homomorphism induces a corresponding functor $\iota^{*}$ (called the extension of scalars) from $R$-modules (where $R=\mathbb{C}[y_{1},y_{2}]$) to $R'$-modules (where $R'=\mathbb{C}[x_{1},x_{2}]$). In this case, it simply replaces a free finite-rank module $R^{n}$ by $R'^{n}$, and on module homomorphisms, it acts by replacing $y_{i}$'s in terms of $x_{i}$'s according to the replacement defined by the maps $Y_{i}$. $W$ is in this way replaced by $W'$, and hence $\iota^{*}$ defines a fusion functor. Applied to the identity defect factorisation $I_{W}$ we obtain a factorisation which describes an interface between a Landau-Ginzburg model with superpotential $W$ (the Kazama-Suzuki model) and a model with superpotential $W'$ (two minimal models).

The fusion of this interface to a defect in the Kazama-Suzuki model is described by the fusion functor $\iota^{*}$. We can also fuse the interface to defects in the minimal models. The action in that direction can indeed be described by a fusion functor $\iota_{*}$ (called restriction of scalars) \cite{Behr:2012xg}. It maps a free finite-rank module $R'^{n}$ to $R^{2n}$, and on a homomorphism (seen as a matrix with polynomial entries in $x_{i}$) it acts by replacing each entry $q (x_{1},x_{2})$ by a $2\times 2$-block,
\begin{equation}\label{iotalowerstar}
\iota_{*}: q (x_{1},x_{2}) \mapsto \begin{pmatrix}
q_{S}\big|_y & (x_{1}-x_{2})q_{A} \big|_{y}\\
\frac{q_{A}}{x_{1}-x_{2}}\big|_{y} & q_{S}\big|_{y}
\end{pmatrix}\ ,
\end{equation} 
where
\begin{equation}
q_S=\tfrac{1}{2}(q(x_1,x_2)+q(x_2,x_2))\quad  ,\quad  q_A=\tfrac{1}{2}(q(x_1,x_2)-q(x_2,x_2))\ .
\end{equation}
The notation $\big|_{y}$ means to express a term in the variables $x_{1},x_{2}$, which is symmetric under the exchange $x_{1}\leftrightarrow x_{2}$, in terms of the variables $y_{i}$. It is straightforward to check that this indeed defines a fusion functor, in particular $W' (x_{1},x_{2})$ is mapped to $W (y_{1},y_{2})\cdot \one_{2}$.

Interfaces between Landau-Ginzburg models that are constructed from a ring homomorphism were investigated in~\cite{Behr:2012xg} (see also~\cite{Dyckerhoff:2011vf,Carqueville:2011zea}). Such \emph{variable transformation interfaces} connect factorisations in different Landau-Ginzburg models whose superpotentials are related by a transformation of variables -- in the example discussed above this provides the opportunity to construct matrix factorisations in the Kazama-Suzuki models by applying the functor $\iota_{*}$ to known factorisations in the minimal models.

\subsection{Fusion functors and rational defects}

We now want to construct fusion functors for rational defects in the $SU (3)/U (2)$ Kazama-Suzuki model. We start with the symmetry defects $D_{((0,0),0;0,\mu)}$. The $\mathbb{Z}_{k+3}$ symmetry that is generated by them is directly connected to the $\mathbb{Z}_{k+3}$ symmetry generated by the ring automorphism
\begin{equation}
\gamma :\mathbb{C}[y_{1},y_{2}]\to \mathbb{C}[y_{1},y_{2}] \quad ,\quad \gamma (y_{1})= \eta^{2}\,y_{1}\quad ,\quad \gamma (y_{2}) = \eta^{4}\, y_{2} \ ,
\end{equation}
where $\eta =e^{\pi i/ (k+3)}$.
Successive application of $\gamma$ leads to a $\mathbb{Z}_{k+3}$-group of automorphisms which leave the superpotential invariant, $\gamma (W (y_{1},y_{2}))=W (y_{1},y_{2})$. Each automorphism $\gamma^{m}$ then induces a fusion functor $(\gamma^{m})^{*}$ that describes the fusion of the defect $D_{((0,0),0;0,6m)}$.

In the minimal models, described by $W' (x_{1},x_{2})=x_{1}^{k+3}+x_{2}^{k+3}$, we also have a class of symmetry defects that are related to the automorphisms
\begin{equation}\label{defofgamma}
\gamma^{(m,n)}:\mathbb{C}[x_{1},x_{2}] \to \mathbb{C}[x_{1},x_{2}]: \begin{cases}
x_1&\mapsto x_1 \eta^{2m}\\
x_2&\mapsto x_2 \eta^{2n}
\end{cases}\ .
\end{equation}
They induce a family of fusion functors $(\gamma^{(m,n)})^{*}$.

We can now obtain defects in the Kazama-Suzuki models by sandwiching defects in the minimal models between the variable transformation interfaces that we introduced before. On the level of fusion functors, this corresponds to applying first $\iota^{*}$, then some fusion functor $U_{\text{mm}}$  describing a minimal model defect, and finally $\iota_{*}$,
\begin{equation}\label{KSdefectsfrommm}
U_{\text{KS}} = \iota_{*} \circ U_{\text{mm}} \circ \iota^{*} \ .
\end{equation}
For example, when we apply this procedure to $(\gamma^{(m,m)})^{*}$ we obtain
\begin{equation}\label{gammamm_sandwiched}
\iota_{*} \circ (\gamma^{(m,m)})^{*} \circ \iota^{*} \cong (\gamma^{m})^{*} \oplus (\gamma^{m})^{*}\ .
\end{equation}
This can be seen as follows. The entry of a homomorphism between $\mathbb{C}[y_{1},y_{2}]$-modules is some polynomial $p (y_{1},y_{2})$. The effect of $\iota^{*}$ is to express this in terms of the variables $x_{i}$, or in other words to map this entry to $p (Y_{1} (x_{1},x_{2}),Y_{2} (x_{1},x_{2}))$. $(\gamma^{(m,m)})^{*}$ then maps this to 
\begin{equation}
p (Y_{1} (\eta^{2m}x_{1},\eta^{2m}x_{2}),Y_{2} (\eta^{2m}x_{1},\eta^{2m}x_{2})) = p (\eta^{2m}Y_{1} (x_{1},x_{2}),\eta^{4m}Y_{2} (x_{1},x_{2})) \ .
\end{equation}
Finally, $\iota_{*}$ maps this to a $2\times 2$-block, but as the expression is already symmetric in $x_{1},x_{2}$, the off-diagonal entries are zero, and we obtain
\begin{equation}
\iota_{*} \circ (\gamma^{(m,m)})^{*} \circ \iota^{*} \big(p (y_{1},y_{2}) \big) = p (\eta^{2m}y_{1},\eta^{4m}y_{2}) \cdot \one_{2}\ .
\end{equation}
This confirms the result in~\eqref{gammamm_sandwiched}.

Instead of starting with $(\gamma^{(m,m)})^{*}$, where we only found a sum of two copies of $(\gamma^{m})^{*}$, we can also apply the procedure to $(\gamma^{(m,n)})^{*}$ with $m$ different from $n$. As an example we can take $m=0$, $n=1$ and build the corresponding fusion functor in the Kazama-Suzuki model by using~\eqref{KSdefectsfrommm}, 
\begin{equation}\label{ansatz:D1}
D_{(1)} := \iota_{*} \circ (\gamma^{(0,1)})^{*} \circ \iota^{*} \ .
\end{equation}
We conjecture that this functor is the fusion functor that corresponds to the defect $D_{((0,0),0;1,3)}$ in the Kazama-Suzuki model, 
\begin{equation}
D_{(1)} (I_{W}) \cong D_{((0,0),0;1,3)} \ .
\end{equation}
Before we give a justification for our claim, we work out the action of the functor on module homomorphisms. A polynomial entry $p (y_{1},y_{2})$ in a homomorphism is first mapped by $\iota^{*}$ to a (symmetric) polynomial in the variables $x_{1},x_{2}$,
\begin{equation}
\tilde{p} (x_{1},x_{2}) = p (Y_{1} (x_{1},x_{2}),Y_{2} (x_{1},x_{2})) \ .
\end{equation}
Applying $(\gamma^{(0,1)})^{*}$ then results in $\tilde{p} (x_{1},\eta^{2}x_{2})$. Finally, $\iota_{*}$ maps this according to~\eqref{iotalowerstar} to a $2\times 2$-matrix by symmetrising and antisymmetrising $\tilde{p}(x_{1},\eta^{2}x_{2})$, so that we arrive at
\begin{equation}\label{D1onp}
D_{(1)} \big(p (y_{1},y_{2}) \big) = \begin{pmatrix}
\frac{\tilde{p} (x_{1},\eta^{2}x_{2})+ \tilde{p} (\eta^{2}x_{1},x_{2})}{2}
& \frac{(x_{1}-x_{2})(\tilde{p} (x_{1},\eta^{2}x_{2})- \tilde{p} (\eta^{2}x_{1},x_{2}) )}{2}\\[2mm]
\frac{\tilde{p} (x_{1},\eta^{2}x_{2})- \tilde{p} (\eta^{2}x_{1},x_{2})}{2 (x_{1}-x_{2})}&
\frac{\tilde{p} (x_{1},\eta^{2}x_{2})+ \tilde{p} (\eta^{2}x_{1},x_{2})}{2}
\end{pmatrix}\Bigg|_{y}\ .
\end{equation}
Note that we used that $\tilde{p}$ is symmetric in its entries.
In~\cite{Behr:2014bta} we checked that this functor applied to matrix factorisations that correspond to some elementary rational boundary conditions gives the result that we expect from the identification with $D_{((0,0),0;1,3)}$ (according to~\eqref{Actionofdefectonbs}). A further check that we want to perform here, is to verify the fusion rules for the rational defects, in particular 
\begin{equation}
D_{((0,0),0;1,3)} \,D_{((0,0),0;1,3)} = D_{((0,0),0;0,6)} + D_{((0,0),0;2,6)} \ .
\end{equation}
As we have already identified $D_{((0,0),0;0,6)}$ with $\gamma^{*}$, we expect that the functor $(D_{(1)})^{2}$ is equivalent to a direct sum of $\gamma^{*}$ and a yet unidentified fusion functor.

The result of applying $D_{(1)}^{2}$ to a polynomial $p (y_{1},y_{2})$ is obtained by applying $D_{(1)}$ once more to~\eqref{D1onp}:
\begin{equation}
\big( D_{(1)}\big)^{2} \big(p (y_{1},y_{2}) \big) = \begin{pmatrix}
\iota_{*}\Big(\frac{\tilde{p} (x_{1},\eta^{4}x_{2})+ \tilde{p} (\eta^{2}x_{1},\eta^{2}x_{2})}{2}\Big)
& \iota_{*}\Big(\frac{(x_{1}-\eta^{2}x_{2})(\tilde{p} (x_{1},\eta^{4}x_{2})- \tilde{p} (\eta^{2}x_{1},\eta^{2}x_{2}) )}{2}\Big)\\[2mm]
\iota_{*}\Big(\frac{\tilde{p} (x_{1},\eta^{4}x_{2})- \tilde{p} (\eta^{2}x_{1},\eta^{2}x_{2})}{2 (x_{1}-\eta^{2}x_{2})}\Big)&
\iota_{*}\Big(\frac{\tilde{p} (x_{1},\eta^{4}x_{2})+ \tilde{p} (\eta^{2}x_{1},\eta^{2}x_{2})}{2}\Big)
\end{pmatrix}
\end{equation}
We can simplify the result by a similarity transformation of this matrix, which corresponds to a natural isomorphism of the corresponding functor. In a first step we eliminate the upper right block: we define $F:=\iota_{*} (x_{1}-\eta^{2} x_{2})$ and 
\begin{equation}
U_{1} = \begin{pmatrix}
\one_{2} & -F\\ 0 & \one_{2}
\end{pmatrix} \ ,
\end{equation}
and compute
\begin{equation}\label{D1squaredfirsttransform}
U_{1} \, \big(D_{(1)}^{2} \big) \big(p (y_{1},y_{2})\big)\, U_{1}^{-1} =\begin{pmatrix}
\iota_{*}\big(\tilde{p} (\eta^{2}x_{1},\eta^{2}x_{2}) \big) & 0\\
\iota_{*}\Big(\frac{\tilde{p} (x_{1},\eta^{4}x_{2})- \tilde{p} (\eta^{2}x_{1},\eta^{2}x_{2})}{2 (x_{1}-\eta^{2}x_{2})}\Big)& \iota_{*}\big(\tilde{p} (x_{1},\eta^{4}x_{2}) \big)
\end{pmatrix} \ .
\end{equation}
This matrix can be brought to a block decomposed form by performing a similarity transformation with
\begin{equation}\label{defofU2}
U_{2} = \begin{pmatrix}
1 & -\frac{\mu_{2}}{\pi_{2}}y_{1} & 0 & 0\\
0 & 1 & 0 & 0\\
0 & \frac{1}{2\pi_{2}} & 1 & 0\\
0 & 0 & 0 & 1
\end{pmatrix} \ ,
\end{equation}
which only affects the lower left block. Here we introduced the notation
\begin{equation}\label{defofpiandmu}
    \pi_p =\frac{1+\eta^p}{2} \qquad,\qquad \mu_p=\frac{1-\eta^p}{2}\ .
\end{equation}
We find (see appendix~\ref{sec:explicitD1squared} for an explicit computation) 
\begin{align}
&U_{2}\,U_{1} \, \big(D_{(1)}^{2} \big) (p (y_{1},y_{2}))\, U_{1}^{-1}\,U_{2}^{-1} \nonumber\\
&= \begin{pmatrix}
\gamma^{(1,1)^{*}} & 0 & 0 & 0\\
0 & \gamma^{(1,1)^{*}} & 0 & 0 \\
\sigma \circ \gamma^{(0,1)^{*}}\circ \tau \circ \gamma^{(0,1)^{*}} & 0  &\sigma \circ \gamma^{(0,2)^{*}} & (x_{1}-x_{2})^{2}\tau \circ \gamma^{(0,2)^{*}}\\
  \tau \circ \gamma^{(0,1)^{*}}\circ \tau \circ \gamma^{(0,1)^{*}}& 0 &   \tau \circ \gamma^{(0,2)^{*}} & \sigma \circ \gamma^{(0,2)^{*}}
\end{pmatrix}(\tilde{p})\Big|_{y} \ ,
\label{D1squaredsecondtransform}
\end{align}
where we introduced the following linear maps on $\mathbb{C}[x_{1},x_{2}]$:
\begin{equation}\label{defsigmatau}
\sigma :q (x_{1},x_{2}) \mapsto \frac{q (x_{1},x_{2})+q (x_{2},x_{1})}{2} \quad ,\quad \tau :q (x_{1},x_{2}) \mapsto \frac{q (x_{1},x_{2})-q (x_{2},x_{1})}{2 (x_{1}-x_{2})} \ .
\end{equation}
The above shows that the functor $D_{(1)}^{2}$ decomposes as \begin{equation}\label{fusionD1D1}
D_{(1)}^{2} \cong \gamma^{*} \oplus D_{(2)} \ , 
\end{equation}
where $D_{(2)}$ is the functor
\begin{equation}\label{defofD2}
D_{(2)} = \begin{pmatrix}
\gamma^{*} & 0\\
\Psi  & \iota_{*} \circ \gamma^{(0,2)*}\circ  \iota^{*}
\end{pmatrix}
\end{equation}
with 
\begin{equation}
\Psi (p (y_{1},y_{2})) =  \begin{pmatrix}
\sigma \circ \gamma^{(0,1)^{*}}\circ \tau \circ \gamma^{(0,1)^{*}}\\
\tau \circ \gamma^{(0,1)^{*}}\circ \tau \circ \gamma^{(0,1)^{*}}
\end{pmatrix} (\tilde{p} (x_{1},x_{2})) \Big|_y \ .
\end{equation}
As we show in appendix~\ref{sec:D1fusion}, this pattern continues, and for $n\leq k+1$ we find that
\begin{equation}\label{decompD1Dn}
D_{(1)}\circ D_{(n)} \cong \gamma^{*}\circ D_{(n-1)} \oplus D_{(n+1)}\ .
\end{equation}
This is in accordance with the conformal field theory expectation,
\begin{equation}
D_{((0,0),0;1,3)} \,D_{((0,0),0;n,3n)} = D_{((0,0),0;n-1,3(n-1)+6)} + D_{((0,0),0;n+1,3(n+1))} \ .
\end{equation} 
These results give strong indications that $D_{(1)}$ correctly implements the fusion of the topological defect $D_{((0,0),0;1,3)}$.

\subsection{More fusion functors as cones}\label{sec:ffascones}

Up to now we have only discussed defects $D_{((0,0),0;l,m)}$ with trivial numerator labels. We have seen strong indications that all of these defects can be realised as fusion functors in the Landau-Ginzburg description.

In this section we make a proposal for fusion functors that describe the defects $D_{((1,0),0;0,4)}$ and $D_{((0,1),0;0,2)}$. These defects can be obtained by a renormalisation group flow from defects with trivial numerator labels (see~(\ref{defectflowI}),(\ref{eq:flowD01})):
\begin{align}
D_{((0,0),0;1,3)} + D_{((0,0),0;0,6)} \ &\longrightarrow 
\ D_{((1,0),0;0,4)}\label{flowtoD10}\\
D_{((0,0),0;1,3)} + D_{((0,0),0;0,0)} \ &\longrightarrow 
\ D_{((0,1),0;0,2)} \ .
\label{flowtoD01}
\end{align}
Let us start with the second flow. We expect that we can construct the factorisation corresponding to $D_{((0,1),0;0,2)}$ as a cone of the factorisations of $D_{((0,0),0;1,3)}$ and $D_{((0,0),0;0,0)}$. For both of them we have a realisation as fusion functors, and we make an ansatz for a fusion functor $D_{(0,1)}$ following the idea of section~\ref{sec:cone} as
\begin{equation}\label{ansatzD01}
D_{(0,1)} = C(\mathrm{Id},D_{(1)},\Psi_{(0,1)}) = \begin{pmatrix}
\mathrm{Id} & 0\\
\Psi_{(0,1)} & D_{(1)}
\end{pmatrix} \ .
\end{equation}
We now need to have a candidate for the entry $\Psi_{(0,1)}$. To find one, we can use again the relation to the defects in minimal models: $D_{(1)}$ is obtained from the minimal model defect that is described by the fusion functor $\gamma^{(0,1)*}$ (see~\eqref{ansatz:D1}). In the minimal models -- as we discussed in section~\ref{sec:minimalmodels} -- we can build a cone $C(\mathrm{Id},\gamma^{(0,1)*},\tilde{\Psi}_{(0,1)})$ from the identity functor $\mathrm{Id}$ and $\gamma^{(0,1)*}$ by setting (see~\eqref{defPsi})
\begin{equation}
\tilde{\Psi }_{(0,1)}= \frac{1}{x_2} \big( \mathrm{Id}-\gamma^{(0,1)*} \big) \ .
\end{equation}
By sandwiching this cone between the functors $\iota_*$ and $\iota^*$ we obtain the cone
\begin{equation}
\iota_* \circ C(\mathrm{Id},\gamma^{(0,1)*},\tilde{\Psi}_{(0,1)}) \circ \iota^* = C(\mathrm{Id}\oplus \mathrm{Id}, D_{(1)}, \iota_* \circ \tilde{\Psi}_{(0,1)} \circ \iota^* )
\end{equation}
based on $D_{(1)}$ and two copies of the identity functor, because
\begin{equation}
\iota_* \circ \mathrm{Id} \circ \iota^* = \mathrm{Id} \oplus \mathrm{Id} \ .
\end{equation}
By a natural isomorphism (for details see appendix~\ref{sec:conedecomposition}) we can see that this cone functor is isomorphic to a sum of one identity functor and the cone we are after:
\begin{equation}\label{ConeD1IdEquivalence}
C(\mathrm{Id}\oplus \mathrm{Id}, D_{(1)}, \iota_* \circ \tilde{\Psi}_{(0,1)} \circ \iota^* ) \cong  \mathrm{Id} \oplus C(\mathrm{Id}, D_{(1)}, \Psi_{(0,1)} )
\end{equation}
with $\Psi_{(0,1)}$ being the left column of $\iota_* \circ \tilde{\Psi}_{(0,1)} \circ \iota^*$. $\Psi_{(0,1)}$ acts on a polynomial $p(y_1,y_2)$ as 
\begin{equation}
\Psi_{(0,1)} \big( p(y_1 ,y_2) \big) = \begin{pmatrix}
\sigma \Big( \frac{1}{x_2} \big(\tilde{p}(x_1,x_2) -\tilde{p}(x_1,\eta^2 x_2 ) \big)\Big)\\[2mm]
\tau \Big( \frac{1}{x_2} \big(\tilde{p}(x_1,x_2) -\tilde{p}(x_1,\eta^2 x_2 ) \big)\Big)
\end{pmatrix}\Bigg|_y
\end{equation}
where $\tilde{p}(x_1,x_2)=p(x_1+x_2,x_1 x_2)$, and $\sigma$ and $\tau$ have been defined in~\eqref{defsigmatau}. This defines the fusion functor $D_{(0,1)}$ (see~\eqref{ansatzD01}), and our claim is that it implements the fusion of the defect $D_{((0,1),0;0,2)}$. A simple check of this claim that we can perform is to act with the functor on a boundary matrix factorisation (which were identified in~\cite{Behr:2010ug} and~\cite{Behr:2014bta}). The simplest boundary factorisation corresponds to the boundary state $|0,0;0\rangle$ and is given by (see appendix~\ref{sec:boundaryfactorisations})
\begin{equation}
Q_{|0,0;0\rangle} = \begin{pmatrix}
0 & y_1^2 - \frac{(1+\eta)^2}{\eta}y_2\\
\frac{W(y_1,y_2)}{y_1^2 - \frac{(1+\eta)^2}{\eta}y_2} & 0
\end{pmatrix} \ .
\end{equation}
It is enough to consider the action on the entry in the upper right of the matrix, and we get
\begin{multline}
D_{(0,1)} \big( y_1^2 - \tfrac{(1+\eta)^2}{\eta}y_2 \big) \\
= \begin{pmatrix}
y_1^2 - \frac{(1+\eta)^2}{\eta}y_2 & 0 & 0\\
\phantom{,}-\frac{(1-\eta^4)(1-\eta)}{2\eta}y_1 & \frac{1+\eta^4}{2}y_1^2-(1+\eta)(1+\eta^3)y_2 \phantom{,}& \frac{1-\eta^4}{2}y_1(y_1^2-4y_2)\\
-\frac{(1+\eta)(1-\eta^4)}{2\eta} & \frac{1-\eta^4}{2}y_1 & \frac{1+\eta^4}{2}y_1^2-(1+\eta)(1+\eta^3)y_2
\end{pmatrix}\ .
\end{multline}
By elementary invertible row and column transformations, this matrix can be brought to the form
\begin{equation}\label{newboundaryMF}
V_1 \, D_{(0,1)} \big( y_1^2 - \tfrac{(1+\eta)^2}{\eta}y_2 \big)\, V_2 = \begin{pmatrix}
y_1^2-\frac{(1+\eta)^2}{\eta}y_2 & 0 & 0 \\
0 & \big(y_1^2-\frac{(1+\eta)^2}{\eta}y_2\big)\big(y_1^2-\frac{(1+\eta^3)^2}{\eta^3}y_2\big) & 0 \\
0 & 0 & 1
\end{pmatrix} \ ,
\end{equation}
where $V_1$ and $V_2$ are given in appendix~\ref{sec:V1V2}.

Comparing the form~\eqref{newboundaryMF} with the boundary factorisations given in appendix~\ref{sec:boundaryfactorisations} we can directly interpret the resulting boundary matrix factorisation as the factorisation corresponding to a superposition of the boundary states $|0,0;0\rangle$ and $|1,0;0\rangle$ -- this is precisely what we expect: evaluating the action of the defect $D_{((0,1),0;0,2)}$ on $|0,0;0\rangle$ according to~\eqref{Actionofdefectonbs} leads to
\begin{equation}
D_{((0,1),0;0,2)} \,|0,0;0\rangle = |0;0,0\rangle + |1,0;0\rangle \ .
\end{equation}
This constitutes a non-trivial check that the interpretation of the fusion functor is indeed correct.

In a similar manner we can identify a fusion functor for the defect $D_{((1,0),0;0,4)}$. Starting from the flow~\eqref{flowtoD10}, we make the ansatz
\begin{equation}
D_{(1,0)} = C((\gamma^1)^*,D_{(1)},\Psi_{(1,0)}) \ .
\end{equation}
To obtain $\Psi_{(1,0)}$, we consider the cone $C(\gamma^{(1,1)*},\gamma^{(0,1)*},\tilde{\Psi}_{(1,0)})$ of fusion functors in the minimal models with 
\begin{equation}
\tilde{\Psi}_{(1,0)} = \frac{1}{x_1}\big(  \gamma^{(1,1)*} - \gamma^{(0,1)*}\big) \ .
\end{equation}
Analogously to the construction of the functor $D_{(0,1)}$ that we discussed before, we define $\Psi_{(1,0)}$ to be the left column of $\iota_*\circ \tilde{\Psi}_{(1,0)}\circ \iota^*$, so it acts on a polynomial $p(y_1,y_2)$ as
\begin{equation}
\Psi_{(1,0)} \big(p(y_1,y_2)\big)= \begin{pmatrix}
\sigma \Big( \frac{1}{x_1} \big(\tilde{p}(\eta^2 x_1,\eta ^2 x_2) -\tilde{p}(x_1,\eta^2 x_2 ) \big)\Big)\\[2mm]
\tau \Big( \frac{1}{x_1} \big(\tilde{p}(\eta^2 x_1,\eta^2 x_2) -\tilde{p}(x_1,\eta^2 x_2 ) \big)\Big)
\end{pmatrix} \Bigg|_y \ ,
\end{equation}
where as before $\tilde{p}(x_1,x_2)=p(x_1+x_2,x_1 x_2)$. Also in this case one can verify that the functor correctly acts on the factorisation corresponding to $|0,0\rangle$.

We have performed a number of additional tests of the identification of the functors $D_{(0,1)}$ and $D_{(1,0)}$ for $D_{((0,1),0;0,2)}$ and $D_{((1,0),0;0,4)}$, respectively. In particular we have shown that
\begin{equation}\label{fusionD01D01}
D_{(0,1)}\circ D_{(0,1)} \cong  D_{(1,0)} \oplus D_{(0,2)} \ ,
\end{equation}
where $D_{(0,2)}$ is a new functor that occurs in this decomposition. This matches precisely the fusion of the corresponding defects (see~\eqref{rationalfusion}),
\begin{equation}
D_{((0,1),0;0,2)}\, D_{((0,1),0;0,2)} = D_{((1,0),0;0,4)} \oplus D_{((0,2),0;0,4)} \ .
\end{equation}
Furthermore, from the flow rule~\eqref{eq:flowD02} we know that $D_{((0,2),0;0,4)}$ arises from a flow starting from a superposition of $D_{((0,0),0;2,6)}$, $D_{((0,0),0;1,3)}$ and $D_{((0,0),0;0,0)}$, and we have shown that the functor $D_{(0,2)}$ can indeed be expressed in the form of successive cones built from $D_{(2)}$, $D_{(1)}$ and $\mathrm{Id}$.

Analogously, the composition of $D_{(1,0)}$ with itself leads to
\begin{equation}\label{fusionD10D10}
D_{(1,0)}\circ D_{(1,0)} \cong D_{(0,1)}\circ \gamma^* \oplus D_{(2,0)}
\end{equation}
which is in agreement with
\begin{equation}
D_{((1,0),0;0,4)}\,D_{((1,0),0;0,4)} = D_{((0,1),0;0,8)} \oplus D_{((2,0),0;0,8)} \ .
\end{equation}
One can show that the functor $D_{(2,0)}$ can be constructed from $(\gamma^2)^*$, $D_{(1)}\circ \gamma^*$ and $D_{(2)}$ by taking successive cones, which matches the flow~\eqref{eq:flowD20} leading to $D_{((2,0),0;0,8)}$.

We have also shown that
\begin{equation}\label{fusionD01D10}
D_{(0,1)}\circ D_{(1,0)} \cong (\gamma^{1})^* \oplus D_{(1,1)} \ ,
\end{equation}
i.e.\ the composition of $D_{(0,1)}$ and $D_{(1,0)}$ can be decomposed as a direct sum of the known functor $(\gamma^1)^*$ and a new functor $D_{(1,1)}$. This again is precisely in accordance with the fusion of the corresponding defects,
\begin{equation}
D_{((0,1),0;0,2)} \,D_{((1,0),0;0,4)} = D_{((0,0),0;0,6)} + D_{((1,1),0;0,6)} \ .
\end{equation}
Also in this case, $D_{(1,1)}$ can be written as a cone in complete agreement with the flow rule~\eqref{eq:flowD11}.
\smallskip

Summarising this section, we have constructed fusion functors $D_{(0,1)}$ and $D_{(1,0)}$ by a cone construction that was inspired by renormalisation group flows between defects in the corresponding conformal field theory, and we have presented evidence that these functors describe the fusion of the defects $D_{((0,1),0;0,2)}$ and $D_{((1,0),0;0,4)}$, respectively.

\subsection{Fusion functors for all rational defects}

In this section we briefly explore how much we can say about possible fusion functors for all rational topological defects. We have identified functors for the basic defects, 
\begin{equation}
D_{((0,0),0;0,6)}\ ,\ D_{((0,0),0;1,3)}\ ,\ D_{((0,1),0;0,2)}\ , \ D_{((1,0),0;0,4)} \ ,
\end{equation}
from which -- as discussed in section~\ref{sec:CFT} (see~\eqref{setofgeneratingdefects}) -- all other rational topological defects can be obtained by fusion. To construct fusion functors for all of these defects we therefore have to compose the functors for the generating defects. The only technical difficulty is that the composition typically has to be decomposed into direct sums of functors, and we have to perform this decomposition to extract the fusion functors we are after. We have already seen several examples of this nature in~\eqref{fusionD1D1}, \eqref{fusionD01D01}, \eqref{fusionD10D10}, and~\eqref{fusionD01D10}: we always find the sought-after functor together with known functors. For example we verified that the fusion of $D_{(1)}$ and $D_{(2)}$ can be decomposed as
\begin{equation}
D_{(1)}\circ D_{(2)} \cong D_{(1)}\circ \gamma^* \oplus D_{(3)}
\end{equation}
with a new functor $D_{(3)}$ that should correspond to the defect $D_{((0,0),0;3,9)}$.

We expect that in this way one can in principle construct fusion functors for all rational topological defects, although in practise the decomposition can be tedious, and therefore we cannot provide a closed formula as of yet.

\section{Conclusion}\label{sec:conclusion}

In this article we have introduced fusion functors on the category of modules as a tool to efficiently compute the fusion of B-type defects and interfaces in $N=(2,2)$ supersymmetric Landau-Ginzburg models. Basic fusion functors are induced from ring homomorphisms, and one can use a generalisation of the cone construction to construct further fusion functors. We have shown in the example of minimal models and $SU(3)/U(2)$ Kazama-Suzuki models that all rational B-type defects seem to have a representation as fusion functors. 

Having the computation of fusion under control allows for an investigation of the (semi-)ring structure of defects with respect to superposition and fusion. In particular, we know that for rational conformal field theories the rational defects close under fusion and form a sub-ring that is isomorphic to the fusion ring of the rational model.\footnote{One even expects a tensor equivalence of subcategories, see~\cite{Davydov:2014oya} for such an equivalence in the case of minimal models.} Identifying such a rational fusion ring inside the category of defect matrix factorisations gives the possibility to identify structures in the Landau-Ginzburg description that reflect the rationality of the corresponding conformal field theory.\footnote{Another strategy to make this connection is to construct the chiral symmetry algebras of the rational theory in the Landau-Ginzburg model~\cite{Witten:1993jg,Nemeschansky:1994is}.} 

Furthermore, our construction has allowed us to propose concrete fusion functors and thus matrix factorisations for rational topological defects in the $SU(3)/U(2)$ model. This provides another part of the dictionary between Landau-Ginzburg models and rational conformal field theories. The rational theories are isolated points in the moduli space of Landau-Ginzburg models, and having a translation between the different descriptions is useful when one wants to combine their complementary power: in the rational theories one has control over the full model and not only over the topological sector, whereas the Landau-Ginzburg description is better suitable to discuss marginal deformations of the models (which break the rationality)~\cite{Jockers:2007ng}. Having identified some of the rational structures on the Landau-Ginzburg side allows us to investigate what happens to them when they are transported along the moduli space.

The discussion of the analogue of the cone construction suggests to view the maps that we employed there as morphisms between fusion functors. This can be done by introducing a category of fusion functors whose morphism spaces are $\mathbb{Z}$-graded. Degree-0 morphisms are the natural transformations, and degree-1 morphisms can be used to construct analogues of cones. It would be interesting to understand the physical meaning of the higher-degree morphisms. A detailed discussion of these structures is left for a further publication.

To conclude, we wish to highlight that some of the matrix factorisations that we have discussed also play a role in Khovanov-Rozansky link homology~\cite{Khovanov:2004}. The factorisations that occur as building blocks in the Khovanov-Rozansky construction~\cite{mackaay2013slnweb,wu2013colored} can be interpreted as defects in Kazama-Suzuki models. We hope that our work could shed some light on these constructions from a physics point of view (see also~\cite{Gukov:2015gmm}) and that they facilitate explicit computations.

\appendix

\section{Boundary factorisations in the Kazama-Suzuki model}
\label{sec:boundaryfactorisations}

Matrix factorisations that correspond to rational B-type boundary states in the $SU(3)/U(2)$ Kazama-Suzuki model were identified in~\cite{Behr:2010ug,Behr:2014bta}. The simplest factorisations arise for the boundary conditions $|L,0;0\rangle$. Their construction is based on a polynomial factorisation of the superpotential, 
\begin{equation}
    W(y_1,y_2) = \prod_{j=0}^{\lfloor\frac{k+1}{2} \rfloor} \underbrace{\left( y_1^2 - \frac{(1+\eta^{2j+1})^2}{\eta^{2j+1}} y_2 \right)}_{\mathcal{J}_j} \ ,
\end{equation}
by grouping the elementary factors $\mathcal{J}_j$. Explicitly, the factorisation corresponding to $|L,0;0\rangle$ is given by 
\begin{equation}
    Q_{|L,0;0\rangle} = \begin{pmatrix}
    0 & \prod_{j=0}^L \mathcal{J}_j\\
    \frac{W}{\prod_{j=0}^L \mathcal{J}_j} & 0
    \end{pmatrix} \ ,
\end{equation}
which obviously satisfies $Q_{|L,0;0\rangle}^2 = W\cdot \mathbf{1}_2$.

\section{Explicit computations}

\subsection{Properties of $\sigma$ and $\tau$}\label{sec:propertiestausigma}

The symmetrisation map $\sigma$ and the antisymmetrisation map $\tau$ (defined in~\eqref{defsigmatau}) are not functors, but they satisfy
\begin{align}
    \sigma(p_1\,p_2) &= \sigma(p_1)\,\sigma(p_2) + (x_1 -x_2)^2 \,\tau(p_1)\,\tau(p_2) \\
    \tau(p_1\,p_2) &= \tau(p_1)\,\sigma(p_2) + \sigma(p_1)\,\tau(p_2) 
\end{align}
when acting on products of polynomials $p_i(x_1,x_2)$.

Recall that the functor $\iota_*$ (introduced below equation~\eqref{defofiota}) acts on a polynomial by replacing the variables $y_1,y_2$ by the elementary symmetric polynomials $x_1+x_2$ and $x_1 x_2$; in particular, it always produces symmetric polynomials in $x_1 ,x_2$. When we act on a symmetric polynomial with the phase shifts $\gamma^{(m,n)*}$ (introduced below equation~\eqref{defofgamma}) and then symmetrise or antisymmetrise, we obtain  
\begin{align}
    \sigma \circ \gamma^{(m,n)*}\circ \iota_* &= \frac{1}{2}\big( \gamma^{(m,n)*}+\gamma^{(n,m)*}\big)\circ \iota_*\\
    \tau  \circ \gamma^{(m,n)*}\circ \iota_* &= \frac{1}{2(x_1-x_2)}\big( \gamma^{(m,n)*}-\gamma^{(n,m)*}\big)\circ \iota_* \ ,
\end{align}
and in particular 
\begin{equation}
    \sigma \circ \gamma^{(m,m)*}\circ \iota_* = \gamma^{(m,m)*}\circ \iota_* \qquad ,\qquad 
    \tau  \circ \gamma^{(m,m)*}\circ \iota_* = 0\ .
\end{equation}
Let us also note that the action of the functor $\iota_*$ (see~\eqref{iotalowerstar}) can be expressed in terms of  $\sigma$ and $\tau$ as
\begin{equation}
    \iota_*\big(q(x_1,x_2\big) = \begin{pmatrix}
    \sigma & (x_1-x_2)^2\, \tau\\
    \tau & \sigma 
    \end{pmatrix} \big( q(x_1,x_2)\big)\Big|_{y} \ .
\end{equation}

\subsection{Fusion of $D_{(1)}$ with itself}
\label{sec:explicitD1squared}

Here we derive the result~\eqref{D1squaredsecondtransform}, so we evaluate explicitly the expression
\begin{align}
    \widetilde{D_{(1)}^2} &= U_{2}\,U_{1} \, \big(D_{(1)}^{2} \big)\, U_{1}^{-1}\,U_{2}^{-1} \ .
\end{align}
Let us first rewrite $U_{1} \, \big(D_{(1)}^{2} \big)\,U_{1}^{-1}$ as given in~\eqref{D1squaredfirsttransform} acting on a polynomial $p(y_1,y_2)$ as
\begin{align}
    &U_{1} \, \big(D_{(1)}^{2} \big)(p)\,U_{1}^{-1}\nonumber \\ 
    &= \begin{pmatrix}
    \gamma^{(1,1)*} & 0 & 0 & 0\\
    0 & \gamma^{(1,1)*} & 0 & 0\\
    \sigma\circ\kappa  & (x_1-x_2)^2\,   \tau\circ\kappa 
     & \sigma\circ \gamma^{(0,2)*} &   (x_1-x_2)^2\, \tau  \circ \gamma^{(0,2)*}\\
    \tau\circ\kappa &  \sigma\circ\kappa  & \tau  \circ \gamma^{(0,2)*} & \sigma\circ \gamma^{(0,2)*}
    \end{pmatrix}\circ \iota^*(p)\Big|_{y}\ ,
\end{align}
where
\begin{equation}
    \kappa = \gamma^{(0,1)*}\circ \tau \circ \gamma^{(0,1)*}\ .
\end{equation}
Now we perform a similarity transformation with the matrix $U_2$ given in~\eqref{defofU2}. It is straightforward to see that this transformation only affects the entries in the lower half of the second column at positions $(3,2)$ and $(4,2)$ in the matrix above. For those components of the resulting matrix $\widetilde{D_{(1)}^2}$ we obtain\footnote{The quantities $\pi_p$ and $\mu_p$ are defined in~\eqref{defofpiandmu}.}
\begin{align}
    &\Big(\widetilde{D_{(1)}^2}\Big)_{(3,2)}(p) \nonumber \\
    &= \left[ (x_1-x_2)^2\, \tau\circ\kappa +\frac{\mu_2}{\pi_2}(x_1+x_2) \sigma\circ\kappa - \frac{1}{2\pi_2} \big(\sigma\circ \gamma^{(0,2)*} - \gamma^{(1,1)*} \big)\right] \circ \iota^*(p)\Big|_{y}\ .
    \label{D1squaredtilde32explicit}
\end{align}
We now evaluate the terms in the square bracket, always using that they act on symmetric polynomials in $x_1,x_2$.
For the first term we compute
\begin{align}
    \tau\circ\kappa &=\tau\circ \gamma^{(0,1)*}\circ \tau \circ \gamma^{(0,1)*} \\
    &= \tau\circ\gamma^{(0,1)*}\circ \left( \frac{1}{2(x_1-x_2)}\big(\gamma^{(0,1)*}-\gamma^{(1,0)*}\big)\right)\\
    &= \tau\circ\left( \frac{1}{2(x_1-\eta^2 x_2)}\big(\gamma^{(0,2)*}-\gamma^{(1,1)*}\big)\right)\\
    &= \frac{1}{2(x_1-\eta^2 x_2)(x_2-\eta^2 x_1)}\tau\circ\Big( (x_2-\eta^2 x_1)\big(\gamma^{(0,2)*}-\gamma^{(1,1)*}\big)\Big)\\
    &= \frac{1}{2(x_1-\eta^2 x_2)(x_2-\eta^2 x_1)}
    \Big( \tau(x_2-\eta^2 x_1)\,\sigma\circ\big(\gamma^{(0,2)*}-\gamma^{(1,1)*}\big) \nonumber \\
    & \quad \qquad \qquad \qquad \qquad \qquad \qquad +  \sigma(x_2-\eta^2 x_1)\,\tau\circ\big(\gamma^{(0,2)*}-\gamma^{(1,1)*}\big)\Big)\\
    &= \frac{1}{2(x_1-\eta^2 x_2)(x_2-\eta^2 x_1)}
    \bigg( -\pi_2 \big( \sigma\circ \gamma^{(0,2)*}-\gamma^{(1,1)*}\big)\nonumber\\
    &  \quad \qquad \qquad \qquad \qquad \qquad \qquad
    + \mu_2 (x_1 +x_2)\, \tau\circ\gamma^{(0,2)*}\bigg)\ .
    \label{resulttaukappa}
\end{align}
In this computation we exploited the properties of $\sigma$ and $\tau$ as presented in the previous section~\ref{sec:propertiestausigma}, and in the last step we used that
\begin{equation}
\tau (x_2-\eta^2 x_1) = - \pi_2 \qquad ,\qquad \sigma(x_2-\eta^2 x_1) = \mu_2 (x_1+x_2) \ .
\end{equation}
Similarly we obtain
\begin{align}
    \sigma\circ\kappa &=\sigma \circ \gamma^{(0,1)*}\circ \tau \circ \gamma^{(0,1)*} \\
    &= \frac{1}{2(x_1-\eta^2 x_2)(x_2-\eta^2 x_1)}
    \bigg(\mu_2 (x_1 +x_2)\, \big( \sigma\circ \gamma^{(0,2)*}-\gamma^{(1,1)*}\big) \nonumber\\
    & \quad \qquad \qquad \qquad \qquad \qquad \qquad 
    -\pi_2 (x_1 -x_2)^2 
    \tau\circ\gamma^{(0,2)*}\bigg)\ .\label{resultsigmakappa}
\end{align}
Plugging in the results~\eqref{resulttaukappa} and~\eqref{resultsigmakappa} into~\eqref{D1squaredtilde32explicit}, we observe that the terms containing $\tau\circ\gamma^{(0,2)*} $ directly cancel, and we are left with
\begin{align}
    \Big(\widetilde{D_{(1)}^2}\Big)_{(3,2)}(p) &= \frac{1}{2(x_1-\eta^2 x_2)(x_2-\eta^2 x_1)}
    \bigg(-\pi_2 (x_1 -x_2)^2 + \frac{(\mu_2)^2}{\pi_2}(x_1+x_2)^2\nonumber\\
    &\quad - \frac{1}{\pi_2}(x_1-\eta^2 x_2)(x_2-\eta^2 x_1) \bigg)
    \big( \sigma\circ \gamma^{(0,2)*}-\gamma^{(1,1)*}\big)\circ \iota^*(p)\Big|_{y}\ .\label{D1squaredtilde32explicit2}
\end{align}
Rewriting
\begin{equation}
    (x_1-\eta^2 x_2)(x_2 -\eta^2 x_1) = (\mu_2)^2 (x_1 +x_2)^2 - (\pi_2)^2 (x_1 -x_2)^2
\end{equation}
we see that all terms in~\eqref{D1squaredtilde32explicit2} cancel, and we find 
\begin{equation}
\Big(\widetilde{D_{(1)}^2}\Big)_{(3,2)} = 0\ .
\end{equation}
Analogously we compute the entry in the fourth row and second column,
\begin{align}
    \Big(\widetilde{D_{(1)}^2}\Big)_{(4,2)}(p)&= \left[ \sigma\circ\kappa +\frac{\mu_2}{\pi_2}(x_1+x_2)\, \tau\circ\kappa - \frac{1}{2\pi_2} \tau \circ \gamma^{(0,2)*}
    \right] \circ \iota^*(p)\Big|_{y} = 0\ ,
    \label{D1squaredtilde42explicit}
\end{align}
which can be confirmed by a straightforward computation using the results above.

\subsection{Decomposition of the induced cone between $\mathrm{Id}\oplus \mathrm{Id}$ and $D_{(1)}$}
\label{sec:conedecomposition}

The cone in~\eqref{ConeD1IdEquivalence}, acting on a polynomial $p(y_1,y_2)$ reads explicitly
\begin{align}
    &C(\mathrm{Id}\oplus \mathrm{Id}, D_{(1)}, \iota_* \circ \tilde{\Psi}_{(0,1)} \circ \iota^* )(p) \nonumber \\
    &= \begin{pmatrix}
    \mathrm{Id} & 0 & 0 & 0\\
    0 & \mathrm{Id} & 0 & 0\\
    \sigma\circ \lambda  & \phantom{,}(x_1-x_2)^2\,\tau\circ \lambda \phantom{,} & \sigma\circ \gamma^{(0,1)*} & \phantom{,}(x_1-x_2)^2\,\tau\circ\gamma^{(0,1)*}\phantom{,}\\
    \tau\circ \lambda  & \sigma\circ \lambda & \tau\circ\gamma^{(0,1)*} & \sigma\circ \gamma^{(0,1)*}
    \end{pmatrix}\circ \iota^* (p) \Big|_y
\end{align}
with
\begin{equation}
    \lambda = \frac{1}{x_2}\big(\mathrm{Id}-\gamma^{(0,1)*}\big) \ .
\end{equation}
When we act with a similarity transformation given by 
\begin{equation}
    U_3 = \begin{pmatrix}
    1 & y_1 & 0 & 0\\
    0 & 1 & 0 & 0\\
    0 & 2 & 0 & 0 \\
    0 & 0 & 0 & 1
    \end{pmatrix} \ ,
\end{equation}
only the third and fourth entry in the second column change, and we obtain
\begin{align}
   &\big( U_3 \,C(\mathrm{Id}\oplus \mathrm{Id}, D_{(1)}, \iota_* \circ \tilde{\Psi}_{(0,1)} \circ \iota^* )\, U_3^-1\big) (p) \nonumber\\
   & = \begin{pmatrix}
    \mathrm{Id} & 0 & 0 & 0\\
    0 & \mathrm{Id} & 0 & 0\\
    \sigma\circ \lambda  & \phantom{,} c_{32}\phantom{,} \  & \sigma\circ \gamma^{(0,1)*} & (x_1-x_2)^2\,\tau\circ\gamma^{(0,1)*}\\
    \tau\circ \lambda  & c_{42} & \tau\circ\gamma^{(0,1)*} & \sigma\circ \gamma^{(0,1)*}
    \end{pmatrix}\circ \iota^* (p) \Big|_y 
\end{align}
with
\begin{align}
\label{c32}
    c_{32} &= (x_1-x_2)^2\, \tau\circ \lambda -(x_1+x_2)\,\sigma\circ\lambda -2 \,\sigma\circ \gamma^{(0,1)*} + 2\,\mathrm{Id}\\
    \label{c42}
    c_{42} &= \sigma \circ \lambda - (x_1+x_2)\,\tau \circ\lambda -2 \,\tau\circ\gamma^{(0,1)*} \ .
\end{align}
We evaluate the individual terms. We have (always assuming that these maps act on symmetric polynomials in $x_1,x_2$)
\begin{align}
    \tau\circ \lambda & = \tau\circ \bigg(\frac{1}{x_2}\big(\mathrm{Id}-\gamma^{(0,1)*}\big)\bigg)\\
    &= \frac{1}{x_1x_2} \tau\circ \big(x_1\big(\mathrm{Id}-\gamma^{(0,1)*}\big)\big)\\
    &= \frac{1}{2\,x_1x_2} \bigg(  \big( \mathrm{Id}-\sigma\circ \gamma^{(0,1)*}\big) - (x_1 +x_2) \tau \circ \gamma^{(0,1)*}\bigg)
\end{align}
and analogously
\begin{equation}
    \sigma \circ \lambda  = \frac{1}{2\,x_1x_2} \bigg( (x_1+x_2) \big( \mathrm{Id}-\sigma\circ \gamma^{(0,1)*}\big) - (x_1 -x_2)^2 \tau \circ \gamma^{(0,1)*} \bigg)\ .
\end{equation}
Inserting these expressions in~\eqref{c32}, we obtain
\begin{align}
    c_{32} &= (x_1 -x_2)^2 \frac{1}{2\,x_1 x_2} \bigg(  \big( \mathrm{Id}-\sigma\circ \gamma^{(0,1)*}\big) - (x_1 +x_2) \tau \circ \gamma^{(0,1)*}\bigg)\nonumber \\
    &\quad - (x_1 +x_2) \frac{1}{2\,x_1 x_2} \bigg( (x_1+x_2) \big( \mathrm{Id}-\sigma\circ \gamma^{(0,1)*}\big) - (x_1 -x_2)^2 \tau \circ \gamma^{(0,1)*} \bigg)\nonumber\\
    &\quad + 2 \big( \mathrm{Id}-\sigma\circ \gamma^{(0,1)*}\big) \\
    &= 0 \ .
\end{align}
Similarly, also $c_{42}$ given in~\eqref{c42} vanishes:
\begin{align}
    c_{42} &= \frac{1}{2\,x_1 x_2} \bigg( (x_1+x_2) \big( \mathrm{Id}-\sigma\circ \gamma^{(0,1)*}\big) - (x_1 -x_2)^2 \tau \circ \gamma^{(0,1)*} \bigg)\nonumber\\
    &\quad -(x_1+x_2)\frac{1}{2\,x_1 x_2} \bigg(  \big( \mathrm{Id}-\sigma\circ \gamma^{(0,1)*}\big) - (x_1 +x_2) \tau \circ \gamma^{(0,1)*}\bigg)\nonumber \\
    &\quad - 2 \,\tau\circ\gamma^{(0,1)*} \\
    & = 0\ .
\end{align}

\subsection{Explicit similarity transformations}
\label{sec:V1V2}

The matrices $V_1$ and $V_2$ needed in the transformation~\eqref{newboundaryMF} are given by
\begin{equation}
V_1 = \begin{pmatrix}
0 & \frac{1+\eta}{\eta(1+\eta^3)} & -\frac{1-\eta}{\eta(1+\eta^3)}y_1\\
\frac{1-\eta^4}{2\eta} & -\frac{(1-\eta^4)}{2\eta(1+\eta^3)}y_1 & \frac{1+\eta^4}{2\eta(1+\eta^3)}y_1^2 -\frac{1+\eta}{\eta}y_2 \\
0 & 0 & \frac{1}{1+\eta}
\end{pmatrix}
\end{equation}
and
\begin{equation}
V_2= \begin{pmatrix}
\frac{\eta}{1+\eta}y_1 & \frac{2\eta}{1-\eta^4}y_1^2 - \frac{2(1+\eta^3)^2}{\eta^2(1-\eta^4)}y_2 & -\frac{2\eta}{1-\eta^4}\\
1 & -\frac{1-\eta^3}{\eta^3}y_1 & 0\\
0 & \frac{1+\eta^3}{\eta^3} & 0
\end{pmatrix} \ .
\end{equation}

\subsection{Successive fusion of $D_{(1)}$}
\label{sec:D1fusion}

Because of the functorial property, a fusion functor in the Kazama-Suzuki model is already determined by its action on $y_1$ and $y_2$. To analyse the successive fusion of $D_{(1)}$ it therefore is sufficient to investigate the effect on the elementary variables, which is given by
\begin{equation}
    D_{(1)}(y_1) = \begin{pmatrix}
    \pi_2\, y_1 & \mu_2 (y_1^2-4\,y_2) \\
    \mu_2 & \pi_2\, y_1
    \end{pmatrix} \ ,\qquad D_{(1)}(y_2) = \eta^2 \,y_2\,\mathbf{1}_2 \ .
\end{equation}
To eliminate the quadratic term in the upper right corner of $D_{(1)}(y_1)$, we perform a similarity transformation and obtain
\begin{equation}
    \widehat{D_{(1)}}(y_1) = \begin{pmatrix}
    1 & -y_1\\ 0 & 1
    \end{pmatrix} D_{(1)}(y_1) \begin{pmatrix}
    1 & -y_1\\ 0 & 1
    \end{pmatrix}^{\!\!\!-1} = \begin{pmatrix}
    \eta^2\,y_1 & -4\,\mu_2\,y_2\\
    \mu_2 & y_1
    \end{pmatrix} \ .
\end{equation}
The similarity transformation acts trivially on $D_{(1)}(y_2)$, and the successive action of $\widehat{D_{(1)}}$ on $y_2$ is given by 
\begin{equation}
    \big(\widehat{D_{(1)}}\big)^n (y_2) = \eta^{2n}\,y_2\,\mathbf{1}_{2^n} \ .
\end{equation}
Hence we can concentrate on the action on $y_1$. Acting twice with $\widehat{D_{(1)}}$ on $y_1$ results in
\begin{equation}
    \big(\widehat{D_{(1)}}\big)^2 (y_1) = 
    \begin{pmatrix}
    \eta^4 \, y_1 & -4\,\eta^2 \,\mu_2\,y_2 & -4\,\eta^2\,\mu_2\,y_2 & 0 \\
    \eta^2\, \mu_2 & \eta^2\, y_1 & 0 & -4\,\eta^2\,\mu_2\,y_2\\
    \mu_2 & 0 &\eta^2\,y_1 & -4\,\mu_2\,y_2 \\
    0 & \mu_2 & \mu_2 & y_1 
    \end{pmatrix} \ .
\end{equation}
We now perform a similarity transformation with
\begin{equation}
    S_1 = \begin{pmatrix}
    1 & 0 & 0 & 0\\
    0 & \frac{1}{1+\eta^2} & -\frac{\eta^2}{1+\eta^2} & 0\\
    0 & 1 & 1 & 0\\
    0 & 0 & 0 & 1
    \end{pmatrix}
\end{equation}
and find (using $(1+\eta^2)\mu_2 = \mu_4$)
\begin{equation}
    S_1\, \big(\widehat{D_{(1)}}\big)^2 (y_1)\,S_1^{-1} = \begin{pmatrix}
    \eta^4\,y_1 & 0 & -4\,\eta^2 \,\mu_2\,y_2 & 0\\
    0 & \eta^2\,y_1 & 0 & 0\\
    \mu_4 & 0 & \eta^2\,y_1 & -4\,\mu_4\,y_2\\
    0 & 0 & \mu_2 & y_1
    \end{pmatrix} \ .
\end{equation}
The action of $\big(\widehat{D_{(1)}}\big)^2$ thus decomposes into the action of $\gamma^*$ and a functor $\widehat{D_{(2)}}$ acting as
\begin{equation}
    \widehat{D_{(2)}} (y_1) = \begin{pmatrix}
    \eta^4\,y_1  & -4\,\eta^2 \,\mu_2\,y_2 & 0\\
    \mu_4  & \eta^2\,y_1 & -4\,\mu_4\,y_2\\
    0  & \mu_2 & y_1
    \end{pmatrix} \ , \qquad \widehat{D_{(2)}} (y_2) =  \eta^{4}\,y_2\,\mathbf{1}_{3} \ .
\end{equation}
This functor is equivalent to the functor $D_{(2)}$ defined in~\eqref{defofD2}, but it is better suited for our purposes here because its action on $y_1$ results in a matrix with entries at most linear in $y_1$ and $y_2$.

When successively applying $\widehat{D_{(1)}}$, the pattern continues, and one finds the decomposition claimed in~\eqref{decompD1Dn} which is realized explicitly as
\begin{equation}\label{D1fusiondecomposition_general}
    \widehat{D_{(1)}}\circ \widehat{D_{(n)}} \cong \gamma^* \circ \widehat{D_{(n-1)}} \oplus \widehat{D_{(n+1)}}
\end{equation}
with $\widehat{D_{(n)}} (y_2) = \eta^{2n}\,y_2\, \mathbf{1}_{n+1}$ and
\begin{equation}
    \widehat{D_{(n)}} (y_1) = \begin{pmatrix}
    \eta^{2n}\,y_1 & -4\eta^{2n-2}\,\mu_2\, y_2 & & & & \\
    \mu_{2n} & \eta^{2n-2}\,y_1 & -4\eta^{2n-4}\,\mu_4\,y_2 & & & \\
     & \mu_{2n-2} & \eta^{2n-4}\,y_1 &  & & \\
      & & & \ddots & & \\
      & & & & \eta^2\,y_1 & -4\mu_{2n}\,y_2\\
       & & & & \mu_2 & y_1
    \end{pmatrix} \  .
\end{equation}
This can be proven by induction over $n$. When we start with the above form of $\widehat{D_{(n)}}(y_1)$ and apply $\widehat{D_{(1)}}$, we obtain
\begin{equation}
    \widehat{D_{(1)}} \big( \widehat{D_{(n)}} (y_1) \big)
  = \scalebox{.75}{\mbox{\ensuremath{\displaystyle\left( \begin{array}{ccccccccc} 
     \cline{1-2} \cline{4-5} 
      \multicolumn{1}{|c}{\eta^{2n+2}y_1} &  \multicolumn{1}{c|}{-4\eta^{2n}\mu_2\,y_2} & & \multicolumn{1}{|c}{-4\eta^{2n}\mu_2\,y_2} & \multicolumn{1}{c|}{0} & & & & \\
     \multicolumn{1}{|c}{\eta^{2n}\mu_{2}} &  \multicolumn{1}{c|}{\eta^{2n}y_1} & & \multicolumn{1}{|c}{0} & \multicolumn{1}{c|}{-4\eta^{2n}\mu_2\,y_2} & & & & \\
  \cline{1-2} \cline{4-5} \\[-4mm]
  \cline{1-2} \cline{4-5} \cline{7-8}  
  \multicolumn{1}{|c}{\mu_{2n}} &  \multicolumn{1}{c|}{0} & & \multicolumn{1}{|c}{\eta^{2n}y_1} &  \multicolumn{1}{c|}{-4\eta^{2n-2}\mu_2\,y_2} & &  \multicolumn{1}{|c}{-4\eta^{2n-2}\mu_4\,y_2} &  \multicolumn{1}{c|}{0} & \\
  \multicolumn{1}{|c}{0} &  \multicolumn{1}{c|}{\mu_{2n}} & & \multicolumn{1}{|c}{\eta^{2n-2}\mu_{2}} &  \multicolumn{1}{c|}{\eta^{2n-2}y_1} & &  \multicolumn{1}{|c}{0} &  \multicolumn{1}{c|}{-4\eta^{2n-2}\mu_4\,y_2} & \\
  \cline{1-2} \cline{4-5} \cline{7-8}  \\[-4mm]
  \cline{4-5} \cline{7-8}
   & & & \multicolumn{1}{|c}{\mu_{2n-2}} &  \multicolumn{1}{c|}{0} & &  \multicolumn{1}{|c}{\eta^{2n-2}y_1} &  \multicolumn{1}{c|}{-4\eta^{2n-4}\mu_2\,y_2} & \\
   & & & \multicolumn{1}{|c}{0} &  \multicolumn{1}{c|}{\mu_{2n-2}} & &  \multicolumn{1}{|c}{\eta^{2n-4}\mu_{2}} &  \multicolumn{1}{c|}{\eta^{2n-4}y_1} & \\
   \cline{4-5} \cline{7-8}
   & & & & & & & &  \ddots 
    \end{array}\right)}}} \, ,
    \end{equation}
where we indicated the blocks that result from the application of $\widehat{D_{(1)}}$ to each matrix element. We can choose another block structure in this matrix, 
\begin{align}
\widehat{D_{(1)}} \big( \widehat{D_{(n)}} (y_1) \big)
  &=
 \scalebox{.72}{\mbox{\ensuremath{\displaystyle \left( \begin{array}{cccccccccc}
\cline{1-1} \cline{3-4}          
\multicolumn{1}{|c|}{\eta^{2n+2}\,y_1} & & \multicolumn{1}{|c}{-4\eta^{2n}\,\mu_2\,y_2} & \multicolumn{1}{c|}{-4\eta^{2n}\,\mu_2\,y_2} & & & & & & \\
\cline{1-1} \cline {3-4}  \\[-4mm]
\cline{1-1} \cline {3-4} \cline {6-7} 
\multicolumn{1}{|c|}{\eta^{2n}\,\mu_{2}} & & \multicolumn{1}{|c}{\eta^{2n}\,y_1} & \multicolumn{1}{c|}{0} & & \multicolumn{1}{|c}{-4\eta^{2n}\,\mu_2\,y_2} &\multicolumn{1}{c|}{0} & & & \\
\multicolumn{1}{|c|}{\mu_{2n}} &  & \multicolumn{1}{|c}{0} & \multicolumn{1}{c|}{\eta^{2n}\,y_1} & &  \multicolumn{1}{|c}{-4\eta^{2n-2}\,\mu_2\,y_2} &  \multicolumn{1}{c|}{-4\eta^{2n-2}\,\mu_4\,y_2} & & & \\
\cline{1-1} \cline{3-4} \cline{6-7} \\[-4mm]
\cline{3-4} \cline{6-7} \cline{9-9}
 & & \multicolumn{1}{|c}{\mu_{2n}} & \multicolumn{1}{c|}{\eta^{2n-2}\,\mu_{2}} & & \multicolumn{1}{|c}{\eta^{2n-2}\,y_1} &  \multicolumn{1}{c|}{0} & &   \multicolumn{1}{|c}{-4\eta^{2n-2}\,\mu_4\,y_2} & \\
 & & \multicolumn{1}{|c}{0} & \multicolumn{1}{c|}{\mu_{2n-2}} & & \multicolumn{1}{|c}{0} & \multicolumn{1}{c|}{\eta^{2n-2}\,y_1} & & \multicolumn{1}{|c}{-4\eta^{2n-4}\,\mu_2\,y_2} & \\
 \cline{3-4} \cline{6-7} \cline{9-9} \\[-4mm]
 \cline{6-7} \cline{9-9} 
 & & & & &  \multicolumn{1}{|c}{\mu_{2n-2}} & \multicolumn{1}{c|}{\eta^{2n-4}\,\mu_{2}} & &  \multicolumn{1}{|c}{\eta^{2n-4}\,y_1} & \\
 & & & & & & & & & \ddots 
    \end{array}\right)}}}\nonumber \\
   & = \begin{pmatrix}
    \eta^{2n+2}y_1 & b_{01} & & & &  \\
    a_{10} & \eta^{2n}y_1\mathbf{1}_2 & B_{12} & & & \\
      & A_{21} & \eta^{2n-2}y_1\mathbf{1}_2 & B_{23} & & \\
       & & \ddots & \ddots & \ddots &  \\
        & & & A_{n\,n-1} & \eta^{2}y_1\mathbf{1}_2 & b_{n\,n+1}\\
        & & & & a_{n+1\,n} & y_1 
    \end{pmatrix}\ ,
\end{align}
with the blocks defined by 
\begin{align}
    A_{i+1\,i} &= \begin{pmatrix}
    \mu_{2n+2-2i} & \eta^{2n-2i}\mu_2 \\
    0 & \mu_{2n-2i}
    \end{pmatrix} &
    B_{i\,i+1} &= -4y_2 \begin{pmatrix}
    \eta^{2n+2-2i}\mu_{2i} & 0\\
    \eta^{2n-2i}\mu_2 & \eta^{2n-2i}\mu_{2i+2}
    \end{pmatrix}\\
    a_{10} &= \begin{pmatrix}
    \eta^{2n}\mu_2\\ \mu_{2n}
    \end{pmatrix} & b_{n\,n+1} &= -4y_2 \begin{pmatrix}
    \eta^2\mu_{2n} \\
    \mu_2
    \end{pmatrix}\\
    a_{n+1\,n} &= \begin{pmatrix}
    \mu_2 & \mu_2
    \end{pmatrix} & 
    b_{01} &= -4y_2 \begin{pmatrix}
    \eta^{2n}\mu_2 & \eta^{2n}\mu_2
    \end{pmatrix}\ .
\end{align}
We now perform a similarity transformation with 
\begin{equation}
    S_n = \begin{pmatrix}
    1 & & & & \\
      & T_1 & & &\\
      & & \ddots & &\\
      & & & T_n & \\
      & & & & 1
    \end{pmatrix}\ ,\quad T_i = \begin{pmatrix}
    \frac{\mu_{2n+2-2i}}{\mu_{2n+2}} & \frac{\mu_{2n+2-2i}}{\mu_{2n+2}} -1 \\
    1 & 1
    \end{pmatrix} \ .
\end{equation}
When we replace $\eta^{2n-2i}\mu_2 =\mu_{2n+2-2i}-\mu_{2n-2i}$ in the blocks $A_{i+1\,i}$, we see that they transform as
\begin{align}
    &T_{i+1}\,A_{i+1\,i}\,T_i^{-1}\nonumber \\
    &\quad = 
    \begin{pmatrix}
    \frac{\mu_{2n-2i}}{\mu_{2n+2}} & \frac{\mu_{2n-2i}}{\mu_{2n+2}} -1 \\
    1 & 1
    \end{pmatrix}\begin{pmatrix}
    \mu_{2n+2-2i} & \mu_{2n+2-2i}-\mu_{2n-2i} \\
    0 & \mu_{2n-2i}
    \end{pmatrix}\begin{pmatrix}
    1 & 1- \frac{\mu_{2n+2-2i}}{\mu_{2n+2}}\\
    -1 & \frac{\mu_{2n+2-2i}}{\mu_{2n+2}}
    \end{pmatrix}\\
    &\quad = \begin{pmatrix}
    \mu_{2n-2i} & 0\\ 
    0 & \mu_{2n+2-2i}
    \end{pmatrix} \ .
\end{align}
We rewrite the blocks $B_{i\,i+1}$ as
\begin{equation}
    B_{i\,i+1} = -4y_2\begin{pmatrix}
    \mu_{2n+2}-\mu_{2n+2-2i} & 0\\\mu_{2n+2-2i} -\mu_{2n-2i} & \mu_{2n+2}-\mu_{2n-2i}
    \end{pmatrix} \ ,
\end{equation}
from where it is straightforward to see that
\begin{align}
    T_i \,B_{i\,i+1}\, T_{i+1}^{-1} &= -4y_2 \begin{pmatrix}
    \mu_{2n+2}-\mu_{2n+2-2i} & 0 \\
    0 & \mu_{2n+2}-\mu_{2n-2i} 
    \end{pmatrix} \\
    &= -4y_2 \begin{pmatrix}
    \eta^{2n+2-2i}\mu_{2i} & 0 \\
    0 & \eta^{2n-2i}\mu_{2i+2}
    \end{pmatrix} \ .
\end{align}
The vector blocks transform into
\begin{align}
    T_1 \,a_{10} &= \begin{pmatrix}
    0\\ \mu_{2n+2}
    \end{pmatrix} & T_n \,b_{n\,n+1} &= 
    \begin{pmatrix}
    0\\ -4\mu_{2n+2}\,y_2
    \end{pmatrix}\\
    a_{n+1\,n}\,T_n^{-1} &= \begin{pmatrix}
    0 & \mu_2 
    \end{pmatrix} & b_{01} \,T_1^{-1} &= \begin{pmatrix}
    0 & -4\eta^{2n}\mu_2\,y_2
    \end{pmatrix}\ .
\end{align}
Putting everything together yields
\begin{align}
&S_n \, \widehat{D_{(1)}} \big( \widehat{D_{(n)}} (y_1) \big) \, S_n^{-1} \\
&\ = \scalebox{.72}{\mbox{\ensuremath{\displaystyle\begin{pmatrix}
\textcolor{blue}{\eta^{2n+2}y_1} & 0 &  \textcolor{blue}{-4\eta^{2n}\mu_2\,y_2} & & & & & & & \\
0 & \textcolor{red}{\eta^{2n}y_1} & 0 & \textcolor{red}{-4\eta^{2n}\mu_2\,y_2} & & & & & & \\
\textcolor{blue}{\mu_{2n+2}} & 0 & \textcolor{blue}{\eta^{2n}y_1} & 0 & \textcolor{blue}{-4\eta^{2n-2}\mu_4\,y_2} & & & & & \\
 & \textcolor{red}{\mu_{2n-2}} & 0 & \textcolor{red}{\eta^{2n-2}y_1} & 0 & \ddots & & & & \\
 & & \textcolor{blue}{\mu_{2n}} & 0  & \textcolor{blue}{\eta^{2n-2}y_1} & \ddots & \textcolor{blue}{-4\eta^{4}\mu_{2n-2}\,y_2} & & &  \\
  & & & \textcolor{red}{\mu_{2n-4}} & 0 & \ddots & 0 & \textcolor{red}{-4\eta^4 \mu_{2n-2}\,y_2} & & \\
  & & & & \textcolor{blue}{\mu_{2n-2}} & \ddots & \textcolor{blue}{\eta^4 y_1} & 0 & \textcolor{blue}{-4\eta^2 \mu_{2n}\,y_2} &  \\
  & & & & & \ddots & 0 & \textcolor{red}{\eta^2 y_1} & 0 & 0\\
  & & & & & & \textcolor{blue}{\mu_4} & 0 & \textcolor{blue}{\eta^2 y_1} & \textcolor{blue}{-4\mu_{2n+2}\,y_2}\\
  & & & & & & & 0 & \textcolor{blue}{\mu_2} & \textcolor{blue}{y_1}
\end{pmatrix}}}}\, .
\end{align}
This decomposes into a block built by the elements in rows and columns $1,3,...,2n+1,2n+2$ (blue color) that we identify with $\widehat{D_{(n+1)}}(y_1)$, and a block built by the elements in rows and columns $2,4,...,2n-2$ (red color) that we identify with $\gamma^* \circ \widehat{D_{(n-1)}} (y_1)$. This proves the decomposition~\eqref{D1fusiondecomposition_general}. It is valid as long as $n\leq k+1$ (for $n=k+2$ we have $\mu_{2n+2}=0$ and the similarity transformations above are ill-defined).  


\end{document}